\documentclass[fleqn,letterpaper,11pt]{article}
\usepackage{graphicx,physics,xcolor,bm,bbm, amsmath,amsfonts,amsbsy,amssymb,fullpage}
\usepackage[numbers,sort&compress]{natbib}
\usepackage[title]{appendix}
\usepackage[hidelinks,colorlinks=true,linkcolor=blue,citecolor=blue,urlcolor=blue]{hyperref}
\usepackage{orcidlink}  

\newcommand{\eq}[2]{\begin{gather} #2  \label{#1} \end{gather}}
\newcommand{\nn}{\nonumber\\}
\begin{document}

\title{\bf Mixed convection and onset of instability due to asymmetric wall heat fluxes in a porous channel}

\author{\bf A. Barletta$^{*}$\orcidlink{0000-0002-6994-5585}, M. Celli\,\orcidlink{0000-0002-2726-4175}, P. V. Brandão\,\orcidlink{0000-0002-8445-3802}}

\date{\small
{\em Department of Industrial Engineering, Alma Mater Studiorum Università di Bologna,\\ Viale Risorgimento 2, 40136 Bologna, Italy}\\[0.4cm]
$^*${corresponding author:\ {\tt antonio.barletta@unibo.it}}
}
\maketitle
%\ead{antonio.barletta@unibo.it}

\begin{abstract}\noindent
The problem of convective instability onset in a horizontal porous channel is explored. The channel's impermeable walls are heated with asymmetric thermal conditions modelled through unequal, but uniform, wall heat fluxes. A stationary solution describing the mixed convection flow is obtained from the governing local balance equations. Then, the linear instability of this flow is analysed by formulating an eigenvalue problem with normal modes. The research specifically highlights the role of the flow rate regime, parametrised through the P\'eclet number, where the Rayleigh number and the heat flux asymmetry ratio are key to defining when instability occurs. The numerical solution of the stability eigenvalue problem is achieved by employing the shooting method. Analytical results are also obtained by employing large-wavelength asymptotic expansions. A numerical analysis is performed to discuss the neutral stability curves and the critical values of the Rayleigh number under different flow and asymmetry conditions. 
~
\\[0.6cm]
\end{abstract}

\section{Introduction}
The study of mixed convection in porous media obtained considerable attention within the thermal science community due to its critical importance in a wide range of engineering and geophysical applications, such as compact heat exchangers, geothermal energy extraction, nuclear waste disposal, building insulation, and subsurface contaminant transport. Mixed convection arises when free convection, governed by the buoyancy force, occurs with an underlying forced flow induced by an externally imposed pressure gradient. The complex interaction between these two modes of heat transfer displays peculiar features in porous media, where the presence of a solid matrix introduces additional resistance to flow and alters the effective thermal diffusivity of the system.

In practical scenarios, porous channels are frequently subject to uniform heat flux boundary conditions \citep{NiBe17}, either due to operational constraints or environmental factors. Such configurations may involve asymmetric wall heat fluxes, where the two confining walls of a porous channel are subjected to unequal thermal input. 
%This asymmetry may affect the otherwise symmetric base flow structure, leading to skewed temperature and velocity fields. 
The resulting velocity and temperature fields reflect the complexity of the flow dynamics, particularly in the transition to buoyancy-induced instability. 

The onset of instability in mixed convection flows through porous channels has been studied under various boundary conditions and flow regimes \citep{prats1966effect, 1010631869741, chung2002onset, ouarzazi2008nonlinear, delache2008weakly, chung2010onset, barletta2012thermal, barletta2014convective, sphaier2014unstable, dubey2018onset, barletta2019unstable, barletta2024linearly, dubey2024influence}, but the influence of asymmetric wall heat fluxes remains an area that is relatively less explored. An exception is the paper by \citet{sphaier2014unstable} which, however, focusses on a quite special configuration with uniform heating from the lower boundary and an adiabatic upper boundary. Instability in such systems can manifest as travelling modes depending on the P\'eclet number. These phenomena not only affect the overall heat transfer performance, but also have implications for system control and stability in real-world applications.

When analysing stability, linear theory is often employed to determine the neutral stability condition and the critical thresholds at which the base flow becomes unstable to small-scale perturbations \citep{straughan2013energy, NiBe17, barletta2019routes}. 
The governing local balance equations are linearised with perturbations superposed to the steady-state base solution.
In the context of a parallel-plate porous channel with asymmetric wall heat fluxes, the base temperature field displays a gradient inclined to the vertical. This, in turn, may result in an unstably stratified fluid density distribution in the vertical direction potentially destabilising the mixed convection flow due to the effect of buoyancy. The degree of wall heat flux asymmetry, characterised by a wall heating ratio, plays a crucial role in modulating the threshold for the onset of the instability.

This study aims to provide a comprehensive analysis of mixed convection and the onset of instability in a horizontal porous channel subject to asymmetric wall heat fluxes. In this approach, the study carried out in this paper provides a development of the analyses presented in \citet{barletta2012thermal, barletta2014convective, barletta2024linearly}, by relaxing the symmetry assumption in the temperature boundary conditions. Furthermore, the very special asymmetric case (uniform heating from the lower boundary with an adiabatic upper boundary) investigated by \citet{sphaier2014unstable} is generalised in the forthcoming analysis. By systematically varying the governing parameters, including 
%the P\'eclet number
{the dimensionless streamwise temperature gradient}, the Rayleigh number and the asymmetry ratio, the conditions leading to the loss of flow stability are identified. The critical thresholds for convective instability are determined using linear stability analysis, and the influence of wall heat flux asymmetry on the structure and nature of the unstable modes is accounted for. The results offer valuable insights into the fundamental mechanisms that drive instability in porous media under non-symmetric thermal boundary conditions, with implications for the design and optimization of thermal systems involving porous structures. An example of such applications is the heat transfer enhancement obtained by using metal foams for the design of innovative heat exchangers \citep{buonomo2020evaluation, bianco2021heat}.

\begin{figure}[t]
\centering
\includegraphics[width=0.8\textwidth]{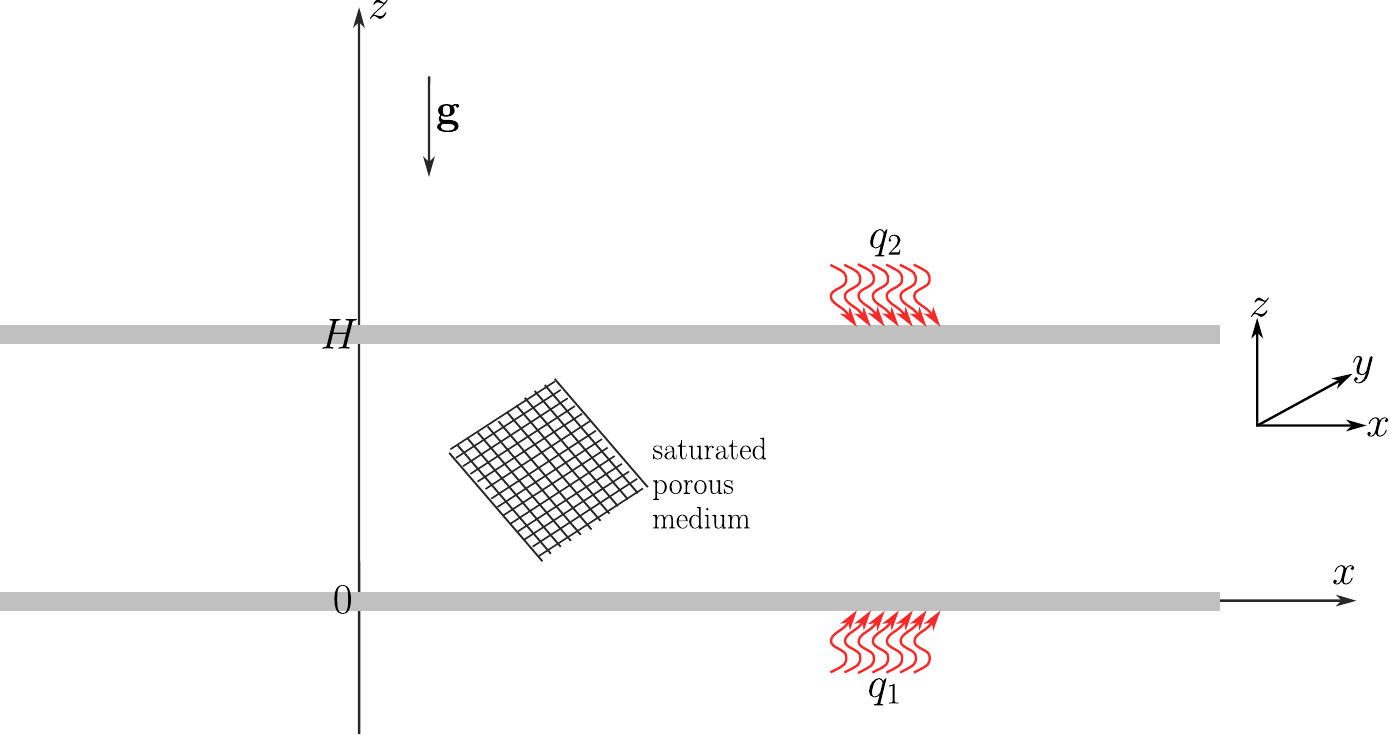}
\caption{\label{fig1}A drawing of the porous channel}
\end{figure}

\section{Mathematical model}
We aim to analyse the buoyant flow in a horizontal porous channel saturated by a fluid. The channel height is $H$ and its width in both the horizontal directions $x$ and $y$ is infinite. The channel impermeable walls,  $z = 0$ and $z=H$, are either heated or cooled by asymmetric and uniform heat fluxes, $q_1$ and $q_2$, which can be positive (incoming flux) or negative (outgoing flux). Figure~\ref{fig1} shows a sketch of the porous channel and of its net wall heating/cooling conditions. 

\subsection{Governing equations}
The buoyant flow is modelled by employing Darcy's law \citep{NiBe17} and the Oberbeck-Boussinesq approximation \cite{rajagopal1996oberbeck, barletta2022boussinesq}. In this framework, the local balance equations for mass, momentum and energy are given by
\eq{1}{
\div{\vb{u}} =0, \nn
 \frac{\mu}{K} \vb{u} = -\grad{p} - \rho_0 \, \beta (T-T_0) \vb{g} , \nn
 \rho_0 \, c\left (\xi \pdv{T}{t}+ \vb{u} \vdot  \grad{T}\right)= \chi \laplacian{T},
}
while the boundary conditions are written as
\eq{2}{
z=0: \qquad  w=0, \quad - \chi \pdv{T}{z}=q_1,\nn
z=H: \qquad w=0, \quad  \chi \pdv{T}{z}=q_2.
}
Here, $\vb{u}$ is the seepage velocity with Cartesian components $\qty(u,v,w)$, $p$ is the local difference between the pressure and the hydrostatic pressure, $T$ is the temperature, $t$ is the time, $\qty(x,y,z)$ are the Cartesian coordinates. Furthermore, $\rho_0$ is the fluid density evaluated at the reference temperature $T_0$, $\mu$ is the dynamic viscosity of the fluid, $K$ is the permeability of the porous medium, $\beta$ is the thermal expansion coefficient of the fluid, $c$ is the specific heat capacity of the fluid, $\xi$ is the ratio between the average volumetric heat capacity of the porous medium and the volumetric heat capacity of the fluid, while $\chi$ is the effective thermal conductivity of the porous medium evaluated as the average value of the conductivities for the solid and fluid phases weighted by the porosity.
The gravitational acceleration is expressed by $\vb{g} = - g\, \vb{e}_z$, where $g$ is its modulus. The unit vectors along the $\qty(x,y,z)$ axes are denoted by $\qty(\vb{e}_x,\vb{e}_y,\vb{e}_z)$.

\subsection{Dimensionless formulation}
The governing equations can be rewritten in a non-dimensional form by means of the scaling
\eq{3}{
\frac{\qty(x,y,z)}{H} \to \qty(x,y,z) \qc \frac{\chi}{ \rho_0 \, c\,  H^2\xi} t \to t \qc  \frac{\rho_0 \, c\, H}{\chi} \vb{u} \to \vb{u} \qc  \frac{\rho_0 \, c\, K}{\chi \, \mu}p\to p, \nn 
\frac{T-T_0}{\Delta T} \to T, \qfor \Delta T=\frac{\chi\, \mu }{\rho_0^2 \, c\, \beta \, g \,K\, H}.
}
Thus, equations \eqref{1} and \eqref{2} can be expressed as
\eq{4}{
\div{\vb{u}} =0, \nn
 \vb{u} = -\grad{p} + T\, \vb{e}_z , \nn
  \pdv{T}{t}+ \vb{u} \vdot  \grad{T}=  \laplacian{T},\nn
  z=0: \qquad w=0, \quad \pdv{T}{z}=-Ra,\nn
  z=1: \qquad w=0, \quad \pdv{T}{z}=\sigma \,Ra,
 }
where $Ra$ is the Darcy--Rayleigh number associated with the heat flux $q_1$ and $\sigma$ is the flux asymmetry ratio,
\eq{5}{
Ra=\frac{\rho_0^2 \, c \, \beta \, g \, K\, H^2  q_1}{\chi^2 \mu }  \qc  \sigma = \frac{q_2}{q_1} .
}
Since $q_1$ and $q_2$ might be either positive or negative, both the Rayleigh number $Ra$ and the asymmetry ratio $\sigma$ can assume a positive or a negative value, accordingly. 

\subsection{Basic state}
A stationary flow solution of \eqref{4} defines the basic state, 
\eq{6}{
\vb{u}_b= \qty[Pe + \gamma\,\qty(\frac{1}{2} - z)] \qty(\cos\varphi\, \vb{e}_x + \sin \varphi\, \vb{e}_y) ,
\nonumber \\   
\grad{T_b} =\gamma\, \qty(\cos\varphi\, \vb{e}_x + \sin \varphi\, \vb{e}_y) + \qty{ Ra\, \qty[(1 + \sigma) z - 1] + \frac{\gamma^2}{2}\, z\,\qty(1-z)} \vb{e}_z   , \nonumber \\
\grad{p}_b = \qty( - u_b, - v_b , T_b)   ,
}
where 
\eq{7}{
\gamma = \frac{Ra (1 + \sigma)}{Pe} ,
}
while the symbol $Pe$ denotes the P\'eclet number relative to the mean flow rate. Thus, the dimensionless mean velocity along the basic flow direction is given by
\eq{8}{
Pe = \int_0^1 \vb{u}_b \vdot \vb{n} \, \dd z , \qfor \vb{n} = \cos\varphi\, \vb{e}_x + \sin \varphi\, \vb{e}_y .
}
A horizontal parallel flow along the $\vb{n}$ direction is implied by \eqref{6} and \eqref{8}. A singularity affects such a flow for $Pe \to 0$, with the only exception of the special case $\sigma = - 1$. In fact, a steady-state flow is allowed only if a net positive or negative heat flux is supplied through the boundaries $z=0,1$ that is convected downstream by the fluid flow or, alternatively, if the net heat flux supplied is zero $(\sigma=-1)$. We mention that this basic state is exactly the same as that considered by Barletta \cite{barletta2012thermal} relative to the symmetric special case $\sigma=1$. On the other hand, the special case $\sigma=-1$ is equivalent to the isoflux-isoflux Horton-Rogers-Lapwood problem contemplated in Table~6.1 of the book by \citet{NiBe17}. We note that the analysis of the latter problem has been recently generalised by \citet{brandao2021stability} and by \citet{turkyilmazoglu2023instability}.

\begin{figure}[t]
\centering
\includegraphics[width=0.8\textwidth]{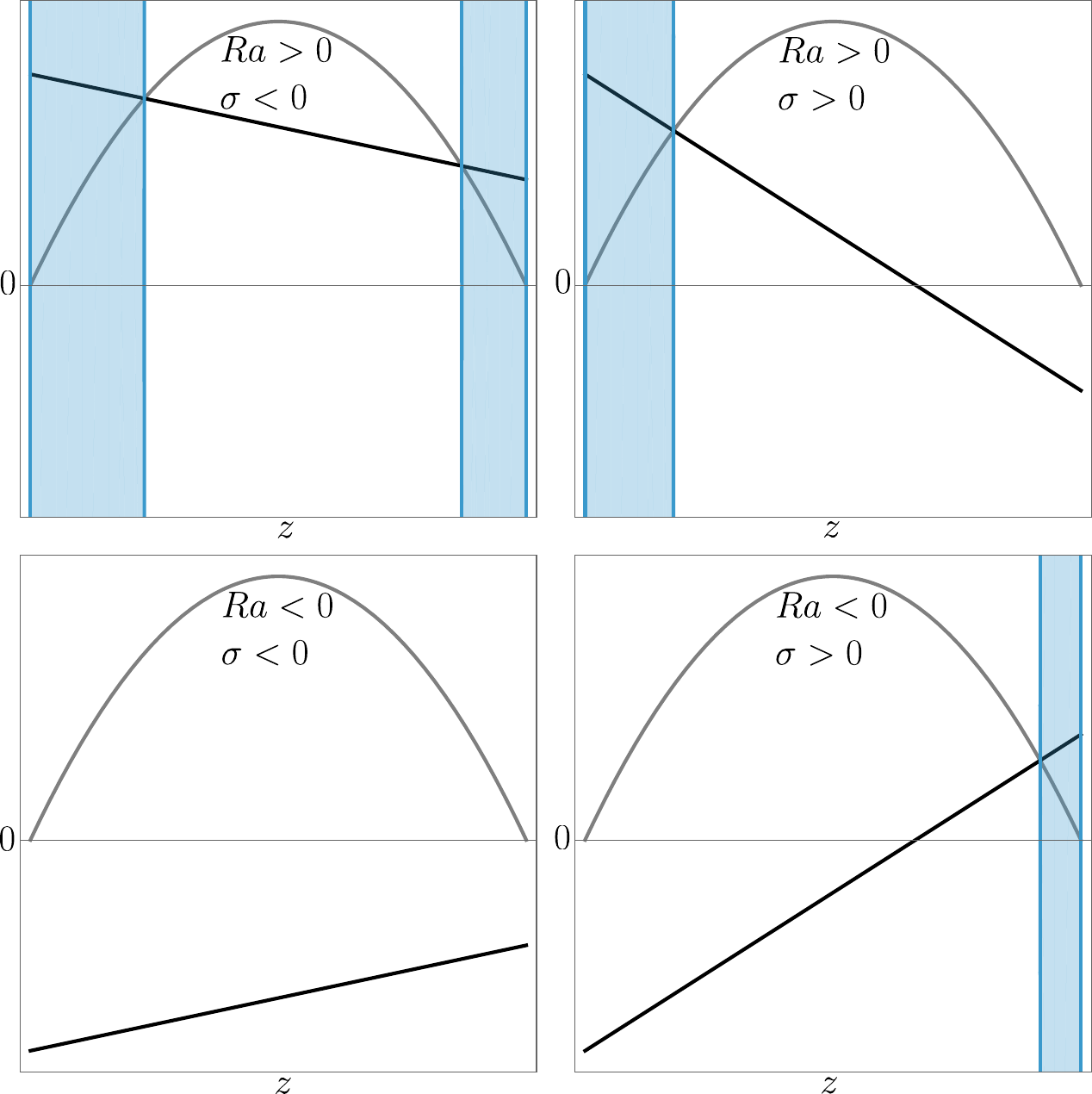}
\caption{\label{fig2}Qualitative plots of the functions of $z$ given by \eqref{9} (grey line) and \eqref{10} (black line); the blue regions denote $\partial T_b/\partial z < 0$}
\end{figure}

\begin{figure}[t]
\centering
\includegraphics[width=0.5\textwidth]{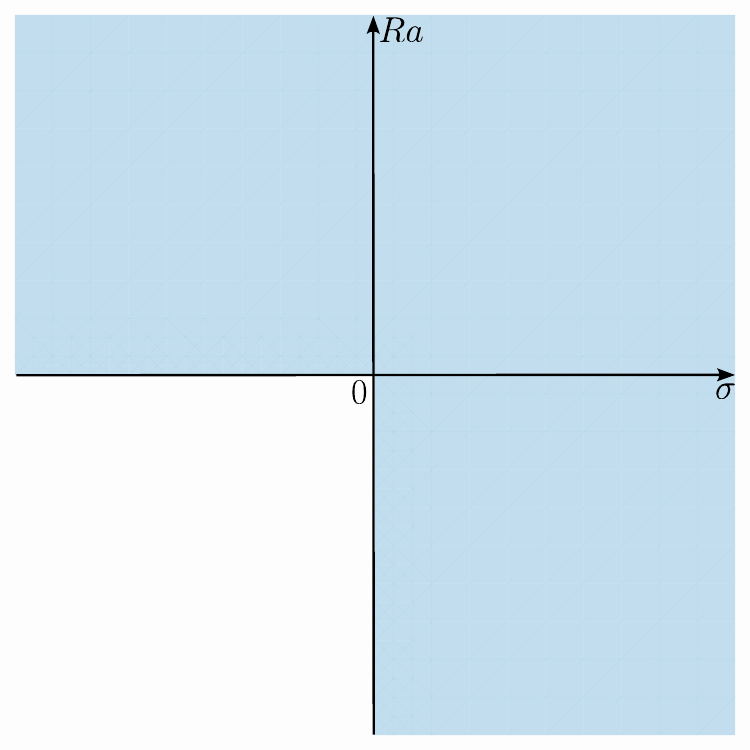}
\caption{\label{fig3}Unstable density stratification: the blue region denotes the parametric conditions for the existence of ranges of $z$ where $\partial T_b/\partial z < 0$}
\end{figure}

One can easily check that \eqref{6} describes a flow state where the vertical component of the temperature gradient can be negative for some ranges of $z$, thus yielding conditions of possibly unstable density stratification. Where and how such ranges of $z$ are localised depend on the three governing parameters $(Ra, Pe, \sigma)$. In fact, a negative $\partial T_b/\partial z$ occurs when
\eq{9}{
\frac{\gamma^2}{2}\, z\,\qty(1-z),
}
is smaller than
\eq{10}{
Ra \qty[1 - (1 + \sigma) z].
}
Such a condition may happen for some subset of the interval $0\le z\le 1$ depending solely on the values of $(Ra, \sigma)$. Figure~\ref{fig2} shows qualitatively how the signs of $Ra$ and $\sigma$ may determine the existence of regions of possibly unstable density stratification, where $\partial T_b/\partial z < 0$. If the existence of regions where $\partial T_b/\partial z < 0$ is determined just by the signs of $Ra$ and $\sigma$ (see Fig.~\ref{fig3}), the value of the parameter $\gamma$ influences the width of such regions.
%
%A case where $\partial T_b/\partial z$ is positive for every $z$ is with $Ra < 0$ and $\sigma < -1$. In such a parametric domain, describing a case of cooling from below and heating from above, no unstable density stratification is possible. We also note that, from the boundary conditions in \eqref{4}, whenever $Ra > 0$, there is always a neighbourhood of the lower wall, $z=0$, where the density stratification is possibly unstable. 
%

Whether a possibly unstable density stratification may yield an actual linear instability can be decided by investigating the dynamics of small perturbations acting on the basic flow state.

\section{Linear instability}
The basic flow state \eqref{6} occurs along a direction $\vb{n}$ inclined an angle $\varphi$ with respect to the $x$ axis. This seemingly useless complication in the mathematics turns out to be, on the contrary, a simplification when the linear stability analysis is formulated. In fact, the general oblique modes of perturbation can be identified by the $y$ independent modes propagating along the $x$ axis. Their directional effect is then easily spanned by letting $\varphi$ vary in the range $\qty[0,\pi/2]$ where, actually, $\varphi=0$ means transverse modes and $\varphi=\pi/2$ longitudinal modes.

\subsection{Streamfunction-temperature formulation}
As already pointed out, the analysis of the onset of linear instability is formulated by employing $y$ independent normal modes, that can be conveniently expressed through a streamfunction, $\psi$, namely
\eq{11}{
 u(x,y,z,t)=u_b(z) + \varepsilon \, \pdv{\psi(x,z,t)}{z} \qc
  v(x,y,z,t)=v_b(z) + \varepsilon \, V(x,z,t), \nn
   w(x,y,z,t)= - \varepsilon \, \pdv{\psi(x,z,t)}{x} \qc
 T(x,y,z,t)=T_b (x,y,z)+ \varepsilon \,  \theta(x,z,t),
}
with $\varepsilon$ the small perturbation parameter $(|\varepsilon| \ll 1)$ and $(\psi, V, \theta)$ the perturbation fields. As a consequence of the streamfunction definition, the local mass balance equation, namely the first equation \eqref{4}, becomes an identity. Furthermore, the $y$ component of the local momentum balance, given by the second equation \eqref{4}, yields $V(x,z,t)=0$.
Eventually, the substitution of \eqref{11} into \eqref{4} yields, to $\order{\varepsilon}$, the linear partial differential equations and boundary conditions for the perturbation fields, with a $(\psi,\theta)$ formulation,
\eq{12}{
\pdv[2]{\psi}{x} + \pdv[2]{\psi}{z} + \pdv{\theta}{x}=0, \nn
\pdv{\theta}{t} +\pdv{T_b}{x} \pdv{\psi}{z}-\pdv{T_b}{z} \pdv{\psi}{x} + u_b\, \pdv{\theta}{x}=\pdv[2]{\theta}{x} +\pdv[2]{\theta}{z} ,\  \ \nn
 z=0,1: \qquad \psi=0 , \quad \pdv{\theta}{z}=0 .
 }

\subsection{Fourier normal modes}
The perturbation fields $(\psi,\theta)$ are now written as Fourier normal modes with wavenumber $k$, time growth rate $\eta$ and angular frequency $\omega$,
%-------------
\eq{13}{
 \psi(x,z,t)= i\,f(z)\, e^{i \, \qty(k x - \omega\, t)}\, e^{\eta \, t} ,\qquad 
  \theta(x,z,t)= h(z)\, e^{i \, \qty(k x - \omega\, t)}\, e^{\eta \, t},
}
%-------------
 where $f(z)$ and $h(z)$ represent the eigenfunctions for the instability eigenvalue problem, formulated by substituting \eqref{13} into \eqref{12}, 
\eq{14}{ 
f'' - k^2\, f+ k\, h=0, \nn
h'' - \qty[ k^2 + i \, k \, \qty(\frac{1}{2} - z) \, \gamma\, \cos\varphi + \eta - i \, \tilde\omega ]\, h 
\nn
\hspace{3cm} -\, i\,\gamma\, \cos\varphi\, f'-  k \,  \qty{ Ra\, \qty[(1 + \sigma) z - 1] + \frac{\gamma^2}{2}\, z\,\qty(1-z)}  \, f = 0 ,\nn
z=0,1: \qquad f=0 , \quad  h'=0 ,
}
where 
\eq{15}{
\tilde\omega = \omega - k\, Pe\, \cos\varphi , 
}
is the reduced angular frequency. 

We note that the eigenvalue problem \eqref{14} depends on the perturbation characteristic parameters $\qty(k, \eta, \tilde\omega)$ and on the flow governing parameters $\qty(Ra, \sigma, \gamma, \varphi)$. If the general case $0 < \varphi < \pi/2$ identifies the oblique modes, the limiting cases $\varphi = 0$ and $\varphi = \pi/2$ define the transverse modes and the longitudinal modes, respectively.

We point out that the eigenvalue problem \eqref{14} loses any explicit dependence on the angle $\varphi$ in the special case where $\gamma \to 0$. This means that, in the limit $\gamma \to 0$, all perturbations ranging from longitudinal to transverse are equivalent in defining the threshold to linear instability. The physical meaning of such a limiting case, for nonzero $Ra$ and $\sigma \ne -1$, is that of an infinitely large P\'eclet number, as one can easily reckon from \eqref{7}. This, in turn, means an extremely large horizontal throughflow across the porous medium or, equivalently, an extremely efficient streamwise convection of the unbalanced wall heat fluxes. As a matter of fact, an independence of the linear instability condition on both $Pe$ and $\varphi$ is exactly what happens also when $\sigma = -1$ (balanced wall heat fluxes), namely, for the isoflux-isoflux Horton-Rogers-Lapwood problem \cite{NiBe17} that, again, means $\gamma = 0$ whatever are the values of $Ra$ and $Pe$.

The numerical solution of \eqref{14} is achieved for input parameters $(k, \sigma, \gamma, \eta, \varphi)$ by determining the eigenvalue pair $(Ra, \tilde\omega)$. In particular, the numerical determination of the neutral stability threshold is achieved by setting $\eta = 0$. The shooting method is employed, according to the general guidelines provided in Chapter 19 of \citet{straughan2013energy} or in Chapter 10 of \citet{barletta2019routes}. The software environment {\sl Wolfram 14} (\copyright{} Wolfram Research, Inc.)~\citep{Mathematica} provides the fundamental  tool for the coding of the solution method, with function {\tt NDSolve} as the numerical solver of the initial value problem starting from $z=0$ and function {\tt FindRoot} for the determination of the parameters fitting the target conditions at $z=1$.

\section{Asymptotic analysis for modes with very large wavelength}
We consider the neutral stability condition, {\em i.e.}, we set $\eta = 0$. We can now express $f(z)$, $h(z)$, $Ra$ and $\tilde\omega$ as power series with respect to $k$ by keeping the lowest orders. In other words, we determine the solution for small wavenumbers $k$, namely for the large wavelength regime. We write
\eq{16}{
f(z) = f_0(z) + f_1(z) k + f_2(z) k^2 + f_3(z) k^3 +\ \ldots\ , \nn
h(z) = h_0(z) + h_1(z) k  + h_2(z) k^2  + h_3(z) k^3 +\ \ldots\ , \nn
Ra = Ra_0  + Ra_1 k + Ra_2 k^2 + Ra_3 k^3 +\ \ldots\ , \nn
\tilde\omega = \tilde\omega_0 + \tilde\omega_1 k  + \tilde\omega_2 k^2  + \tilde\omega_3 k^3 +\ \ldots\ . 
}
By substituting \eqref{16} into \eqref{14} and by collecting like powers of $k$, we obtain to $\order{k^0}$,
\eq{17}{
f''_0 = 0 \qc \qfor f_0(0) =0 \qc f_0(1) =0, \nn
h''_0 - i\,\gamma\, \cos\varphi\, f'_0 + i \, \tilde\omega_0\, h_0 = 0 \qc \qfor h'_0(0) =0 \qc h'_0(1) =0,
}
where the first \eqref{17} yields $f_0(z)=0$ for $0\le z\le 1$, while the second \eqref{17} yields $\tilde\omega_0 = 0$ and $h_0(z) = constant$. Such a constant can be arbitrarily chosen meaning that we can gauge the otherwise arbitrary amplitude of the perturbation. A convenient choice is $h(0) = 1$, so that we can write $h_0(0) = 1$ and $h_n (0) = 0$, for every integer $n>0$. Thus, we obtain $h_0(z) = 1$ for $0\le z\le 1$.
To $\order{k}$, we get
\eq{18}{
f''_1 + 1 = 0 \qc \qfor f_1(0) =0 \qc f_1(1) =0, \nn
h''_1 - i\, \gamma\, \cos\varphi \, f'_1 - i \qty[ \qty(\frac{1}{2} - z)\, \gamma\, \cos\varphi - \tilde\omega_1] = 0 , \nn 
\hspace{3cm}\qfor h_1(0) = 0 \qc h'_1(0) = 0 \qc h'_1(1) = 0 ,
}
namely,
\eq{19}{
f_1(z) = \frac{1}{2} z \left(1-z\right) \qc 
h_1(z) = -\frac{i z^2}{6}  \qty( 2 z -3 ) \, \gamma \, \cos\varphi \qc \tilde\omega_1 = 0 .
}
To $\order{k^2}$, one has
\eq{20}{
f''_2 + h_1 = 0 \qc \qfor f_2(0) =0 \qc f_2(1) =0, \nn
h''_2 - i \, \gamma \cos\varphi \,  f'_2 - i\, \qty( \frac{1}{2} - z) \, \gamma\, \cos\varphi \, h_1 - \qty{ Ra_0\, \qty[ (\sigma + 1)\, z - 1 ] + \frac{\gamma^2}{2}\, z\, \qty( 1 - z ) } \, f_1 \nn
\hspace{3cm} -\, 1 + i\, \tilde\omega_2 = 0  \qc \qfor h_2(0) = 0 \qc h'_2(0) = 0 \qc h'_2(1) = 0 .
}
We do not write here the expressions of $f_2(z)$ and $h_2(z)$ for the sake of brevity. We just say that \eqref{20}  serves to determine explicitly $Ra_0$ and $\tilde\omega_2$,
\eq{21}{
Ra_0 = \frac{120 + \gamma^2 \, \qty( 1+ 2\,\cos^2\!\varphi )}{5\, \qty(1 - \sigma)} \qc \tilde\omega_2 = 0 .
}
A finite $Ra_0$ exists for every $(\gamma, \sigma)$ except for $\sigma = 1$, which is the special case studied in \citet{barletta2012thermal} and extended to the anisotropic case in \citet{barletta2024linearly}.
We also note that $|Ra_0|$ is a monotonic decreasing function of $\varphi$, for prescribed $(\gamma, \sigma)$, meaning that the lowest neutral stability values of $|Ra|$ in the limit $k \to 0$ are obtained with longitudinal modes $(\varphi = \pi/2)$.

\begin{figure}[t]
\centering
\includegraphics[width=0.8\textwidth]{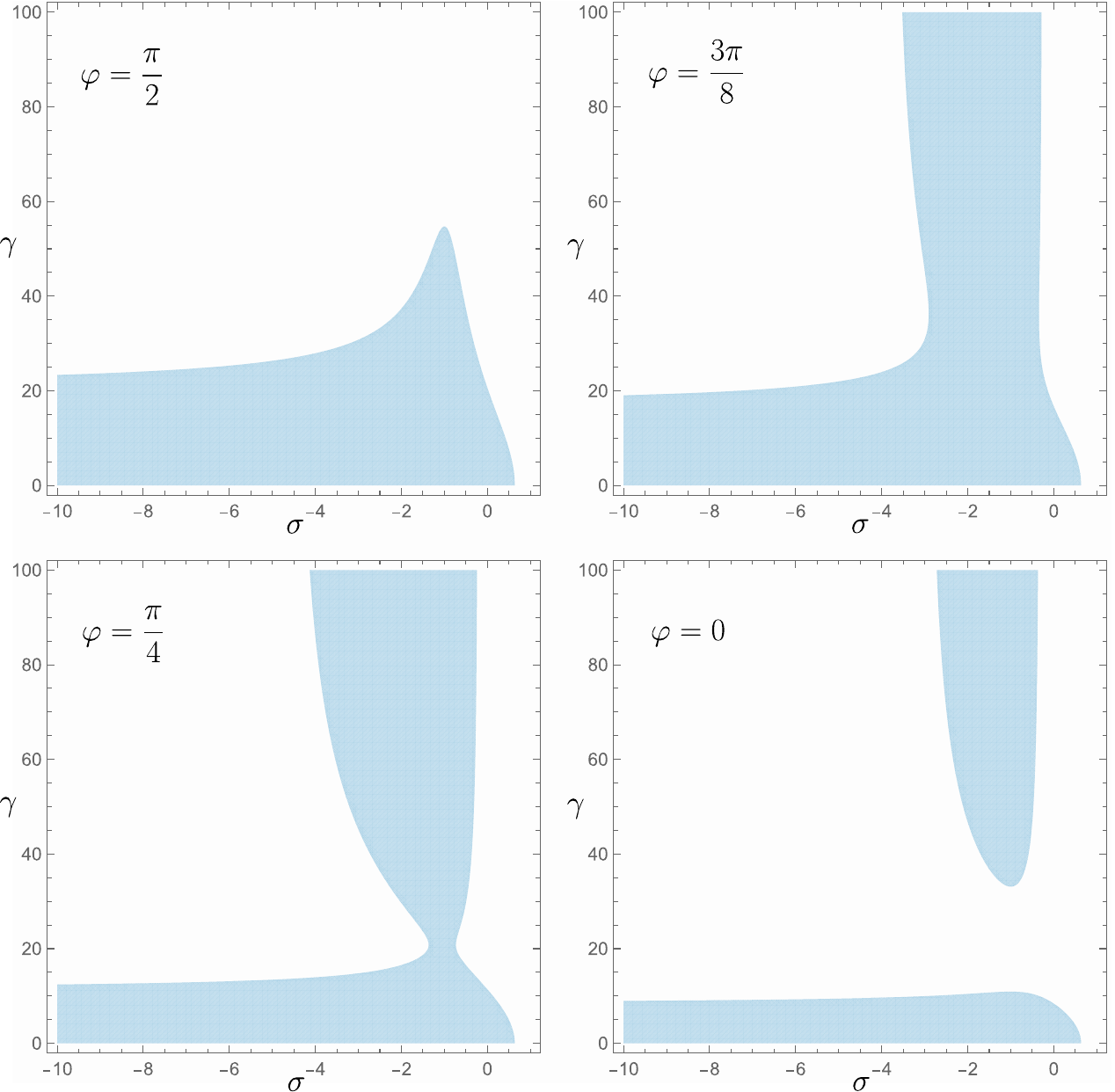}
\caption{\label{fig4}Regions where $Ra_2 > 0$ (shaded in blue), for $\sigma < 1$ and differently oriented normal modes}
\end{figure}

\begin{figure}[t]
\centering
\includegraphics[height=7cm]{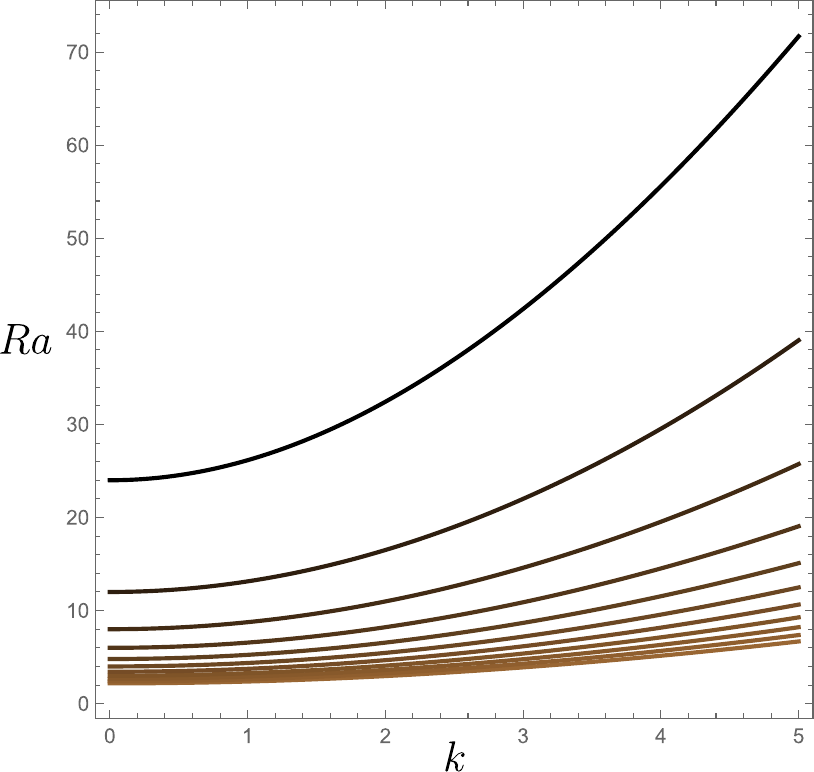}~{\small (a)} \includegraphics[height=7cm]{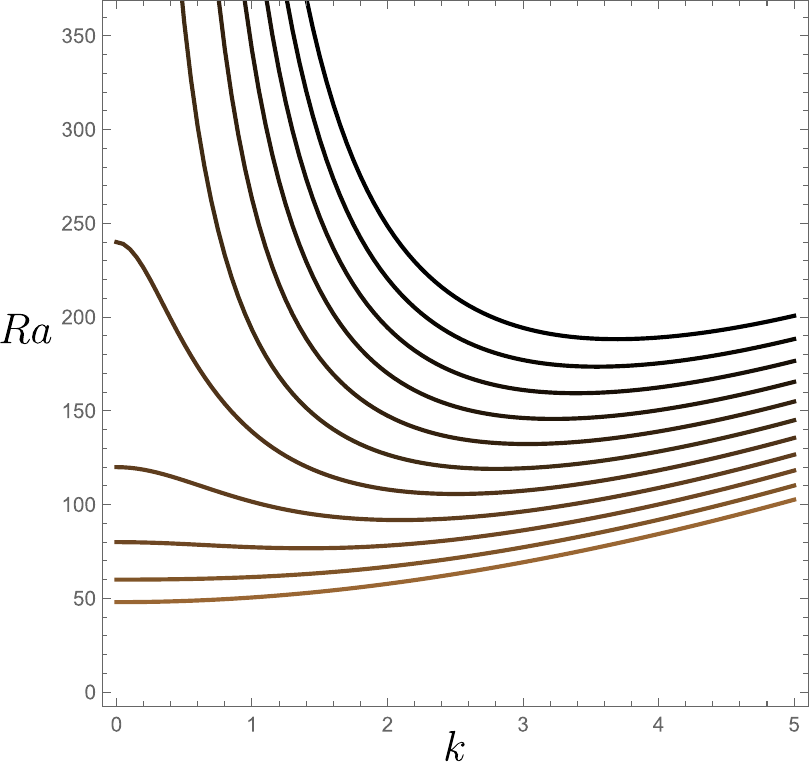}~{\small (b)}
\caption{\label{fig5}Limiting case $\gamma\to 0$, neutral stability curves $Ra(k)$ with: (a) $\sigma$ ranging from $-10$ to $0$ in steps of $1$; (b) $\sigma$ ranging from $0.5$ to $1.5$ in steps of $0.1$. Increasing values of $\sigma$ are identified with a line colour shifting from brown to black}
\end{figure}

\begin{figure}[t]
\centering
\includegraphics[height=7cm]{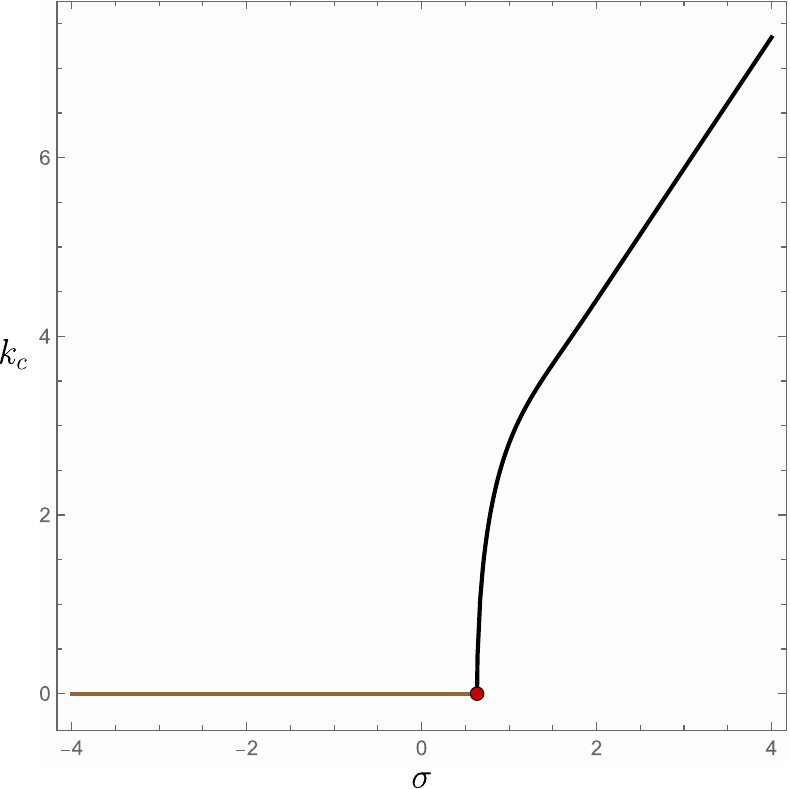}~{\small (a)} \includegraphics[height=7cm]{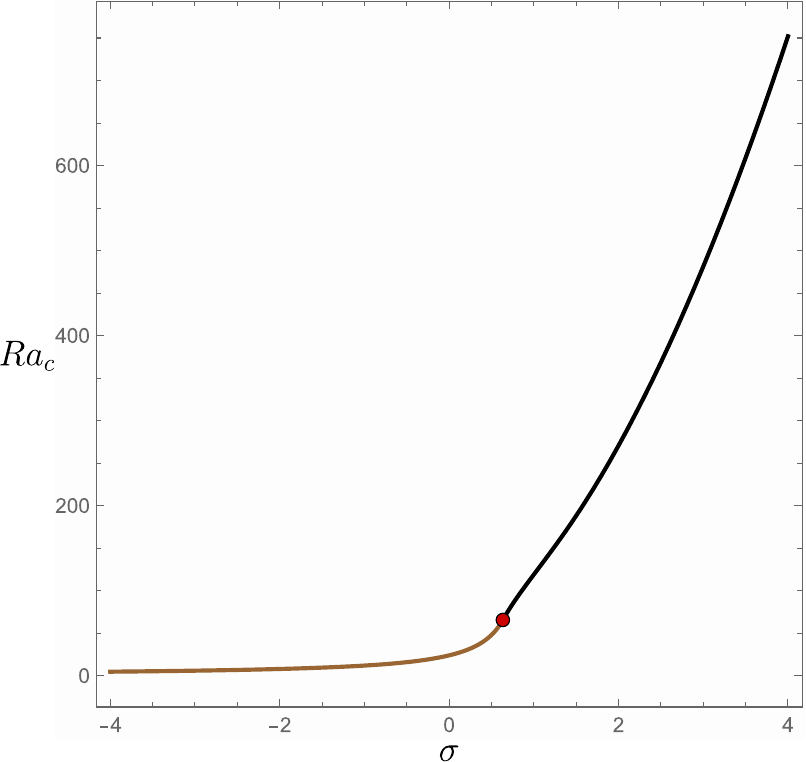}~{\small (b)}
\caption{\label{fig6}Limiting case $\gamma\to 0$, Critical values of $k$ (a) and $Ra$ (b) versus $\sigma$. The brown lines are relative to the range defined by \eqref{27} where $k_c=0$ and $Ra_c = Ra_0$, while the black lines denote cases where $k_c>0$}
\end{figure}

\begin{figure}[t]
\centering
\includegraphics[height=7cm]{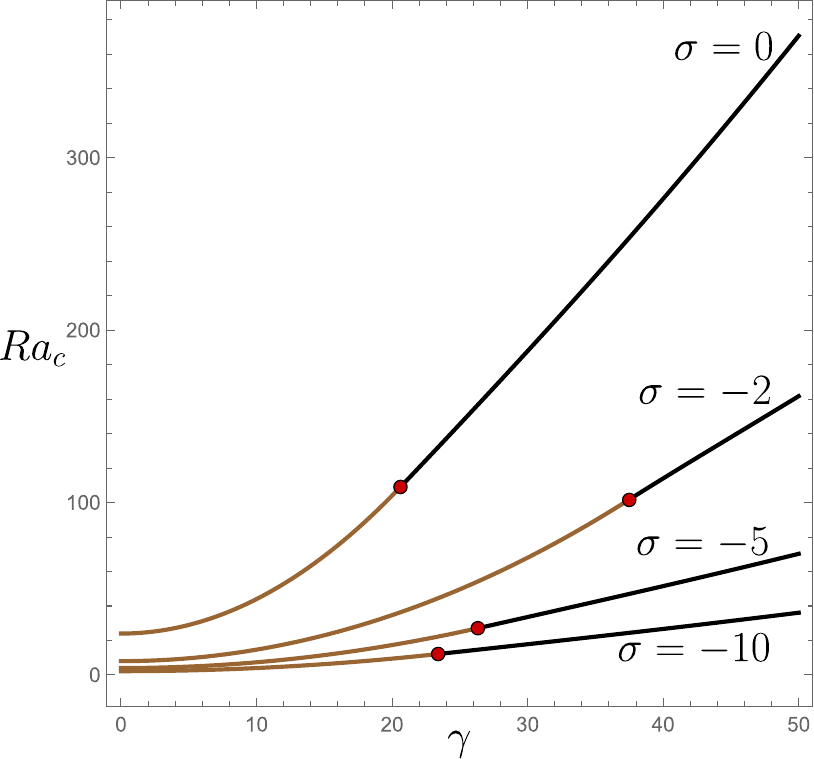}~{\small (a)} \includegraphics[height=7cm]{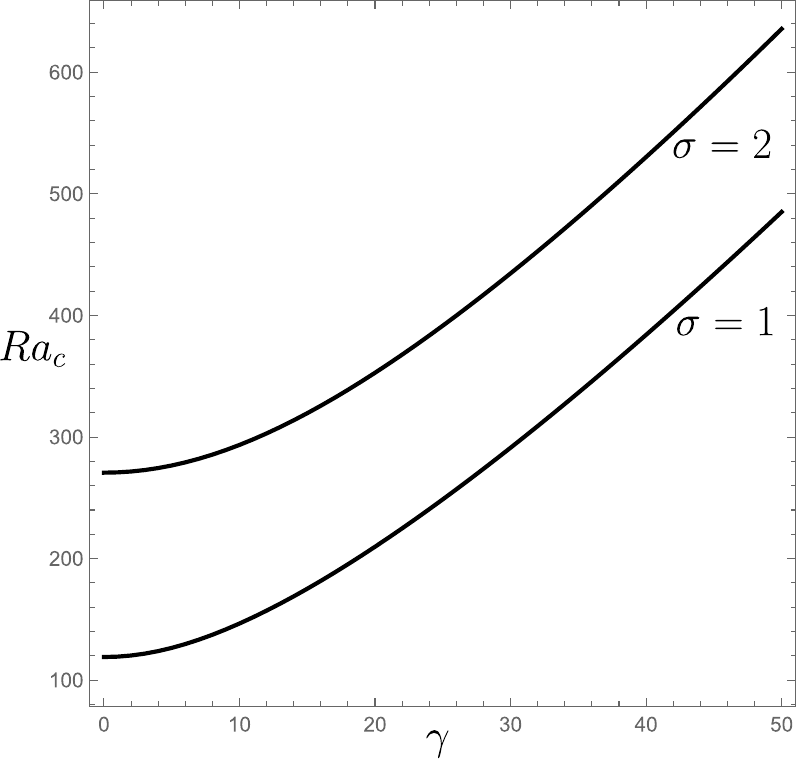}~{\small (b)}
\caption{\label{fig7}Critical values of $Ra$ versus $\gamma$ for different values of $\sigma \le 0$ (a) or $\sigma > 0$ (b), relative to the longitudinal modes $(\varphi = \pi/2)$. The brown lines are for the range where $k_c=0$ and $Ra_c = Ra_0$, while the black lines denote cases where $k_c>0$}
\end{figure}

One may develop also the $\order{k^3}$ and $\order{k^4}$ differential problems. We mention that the $\order{k^3}$ solution provides the evaluation of $Ra_1$ and $\tilde\omega_3$,
\eq{22}{
Ra_1 = 0 \qc \tilde\omega_3 = \frac{ \qty(1 + \sigma ) \qty[ 120 + \gamma ^2 \, \qty( 1 + 2\, \cos^2\!\varphi ) ]}{6720 \qty(1 - \sigma)}\,  \gamma \, \cos\varphi .
}
From \eqref{22}, we understand that oblique and transverse rolls at neutral stability have, in general, a nonzero $\tilde\omega$ when $k \ne 0$. As for the the $\order{k^4}$ solution, one may get expressions for $Ra_2$ and $\tilde\omega_4$,
\eq{23}{
Ra_2 = - \frac{\gamma ^4 \qty(23 \sigma^2 + 42 \sigma + 23) + 880\, \gamma^2 \qty(5 \sigma^2+14 \sigma +5 ) - 316800 (19 \sigma^2 - 42 \sigma + 19 )}{2772000 \qty(1 - \sigma )^3}
\nn
\hspace{2cm}-\frac{\gamma ^2 \left(29 \sigma ^2+206 \sigma +29\right)+7040 \left(5 \sigma ^2-7 \sigma +5\right)}{1386000 (1-\sigma )^3} \, \gamma^2 \cos^2\!\varphi 
\nn
\hspace{4cm}-\frac{137 \sigma ^2+78 \sigma +137 }{5544000 (1-\sigma )^3}\, \gamma^4 \cos(2 \varphi ) \cos^2\!\varphi  
\qc 
\tilde\omega_4 = 0 .
}
The sign of $Ra_2$ is very important in defining whether the neutral stability value of $Ra$ is an increasing or decreasing function of $k$ starting from $0$. If $Ra_0 > 0$ and $Ra_2<0$, one is sure that the minimum of the neutral stability curve is not achieved with $k \to 0$, but with a critical wave number $k = k_c > 0$. As is well known, the neutral stability value of $Ra$ for $k=k_c$ is termed the critical Rayleigh number and is denoted with $Ra_c$.

\section{Symmetries}
On varying the governing parameters $(Ra, \sigma, \gamma)$, one can test the conditions for the onset of linear instability. In the determination of the neutral stability curves in the $(k,Ra)$ plane and the evaluation of the critical values $(k_c, Ra_c, \tilde\omega_c)$, the analysis is grounded on the mathematical symmetries underlying the eigenvalue problem \eqref{14}.

\begin{figure}[ht!]
\centering
\includegraphics[height=7cm]{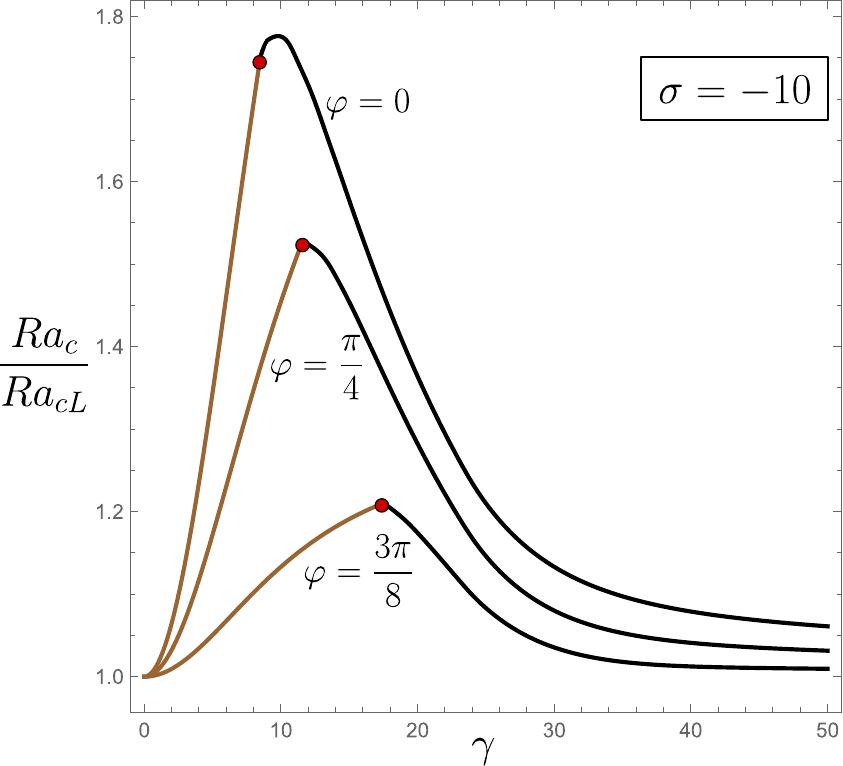}~ \includegraphics[height=7cm]{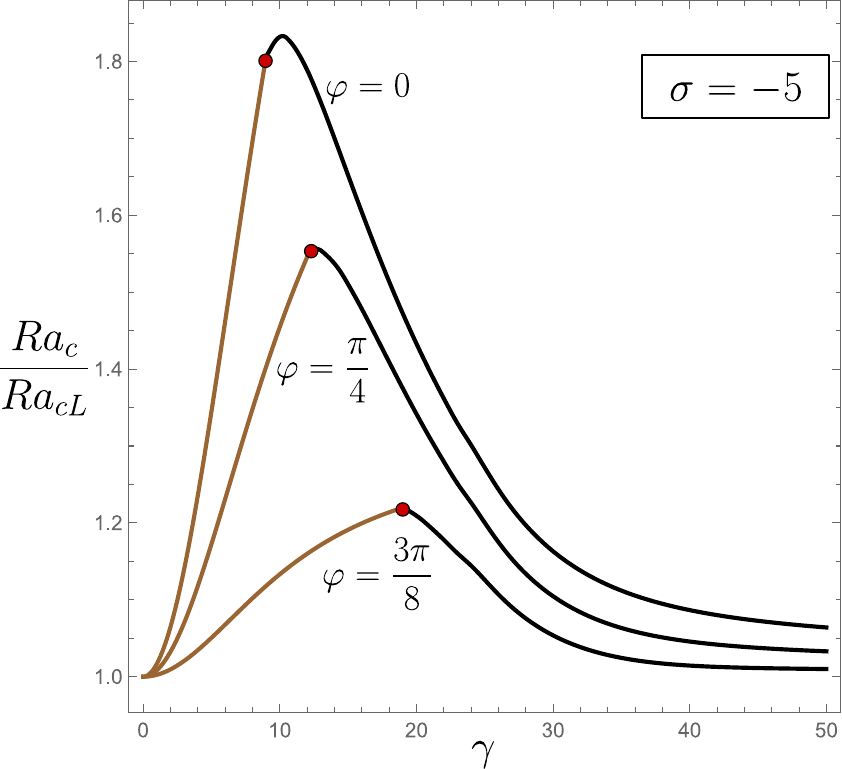}\\[4pt]
\includegraphics[height=7cm]{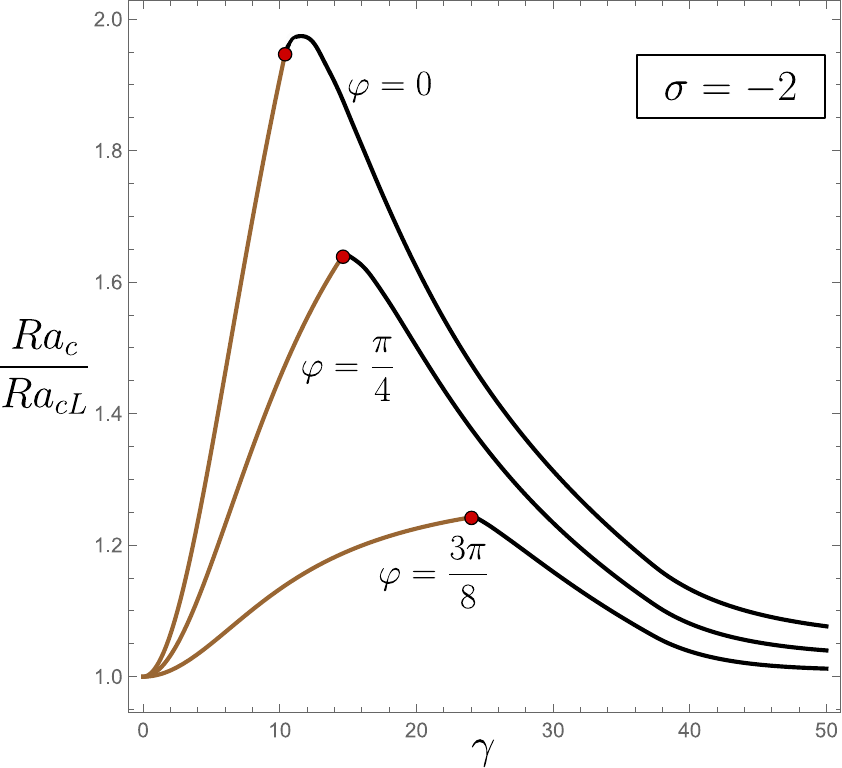}~ \includegraphics[height=7cm]{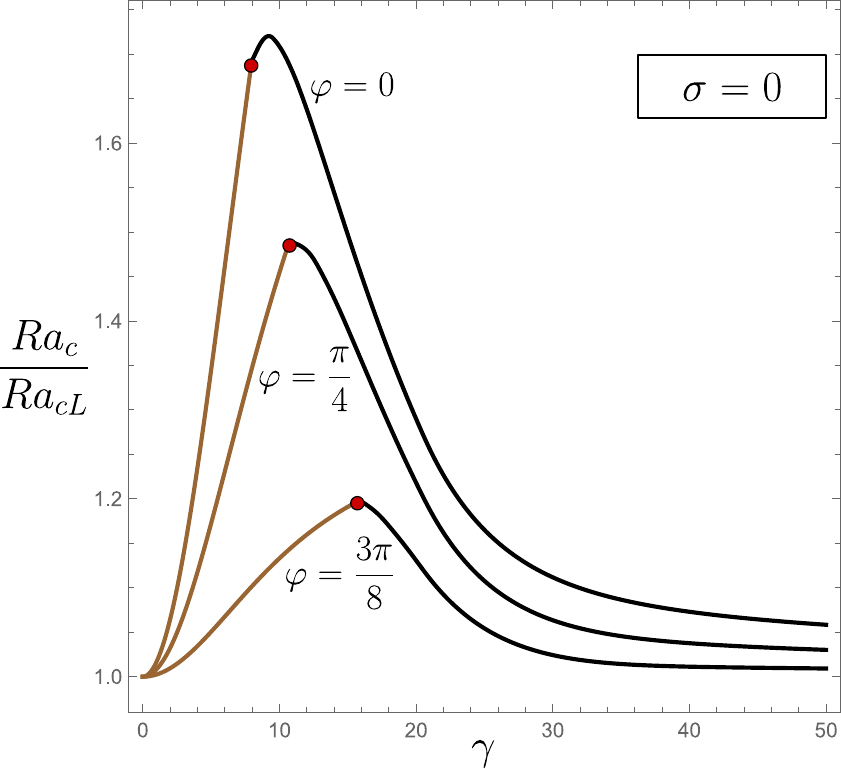}
\caption{\label{fig8}Ratio between the critical values of $Ra$ evaluated for oblique or transverse modes and those for longitudinal $(L)$ modes versus $\gamma$ for different values of $\sigma \le 0$. The brown lines are for the range where $k_c=0$ and $Ra_c = Ra_0$}
\end{figure}

\begin{figure}[ht!]
\centering
\includegraphics[height=7cm]{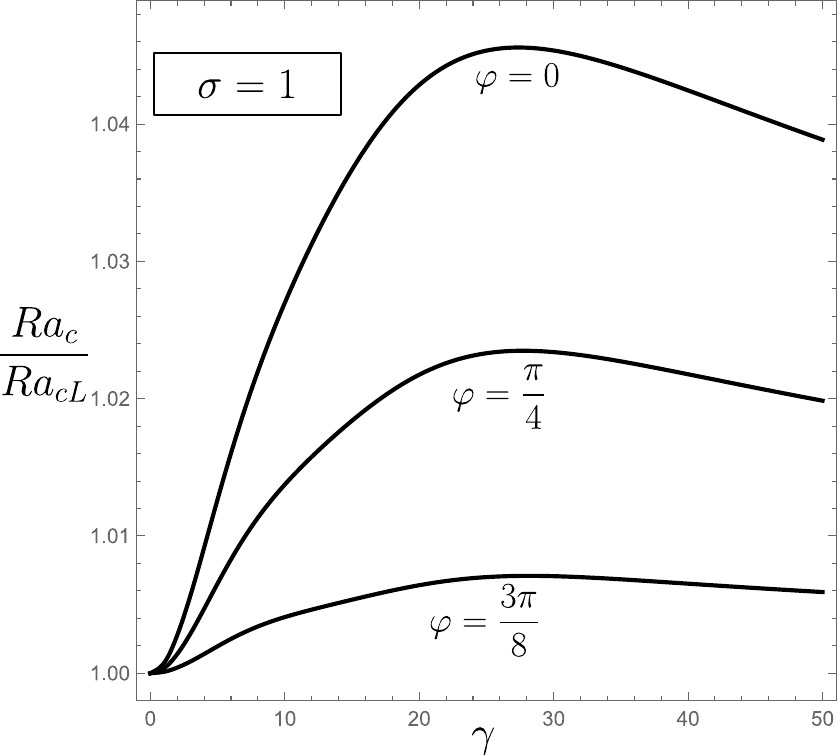}~ \includegraphics[height=7cm]{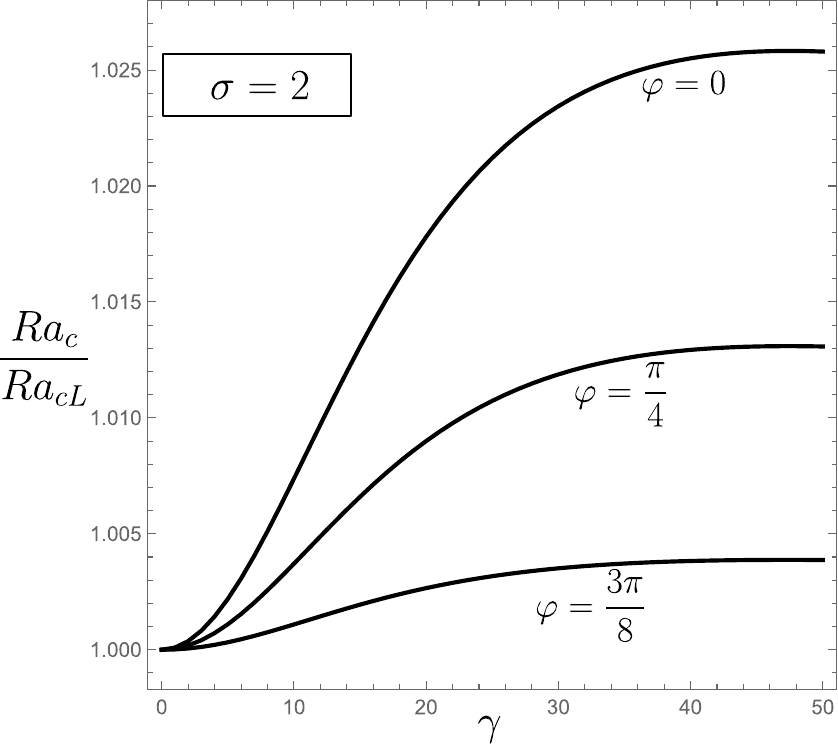}
\caption{\label{fig9}Ratio between the critical values of $Ra$ evaluated for oblique or transverse modes and those for longitudinal $(L)$ modes versus $\gamma$ for different values of $\sigma > 0$}
\end{figure}

\begin{figure}[ht!]
\centering
\includegraphics[height=7cm]{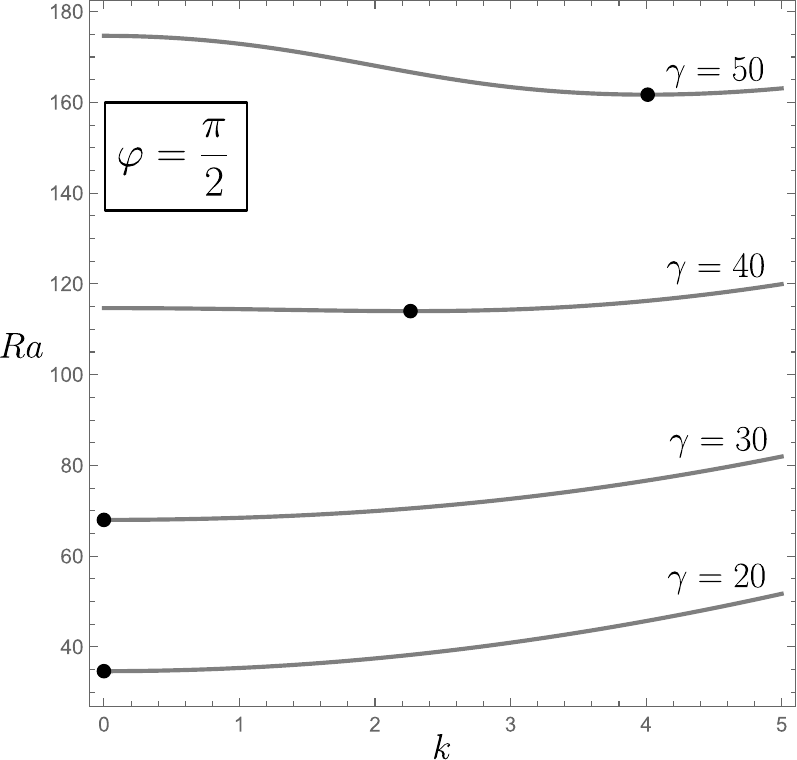}~ \includegraphics[height=7cm]{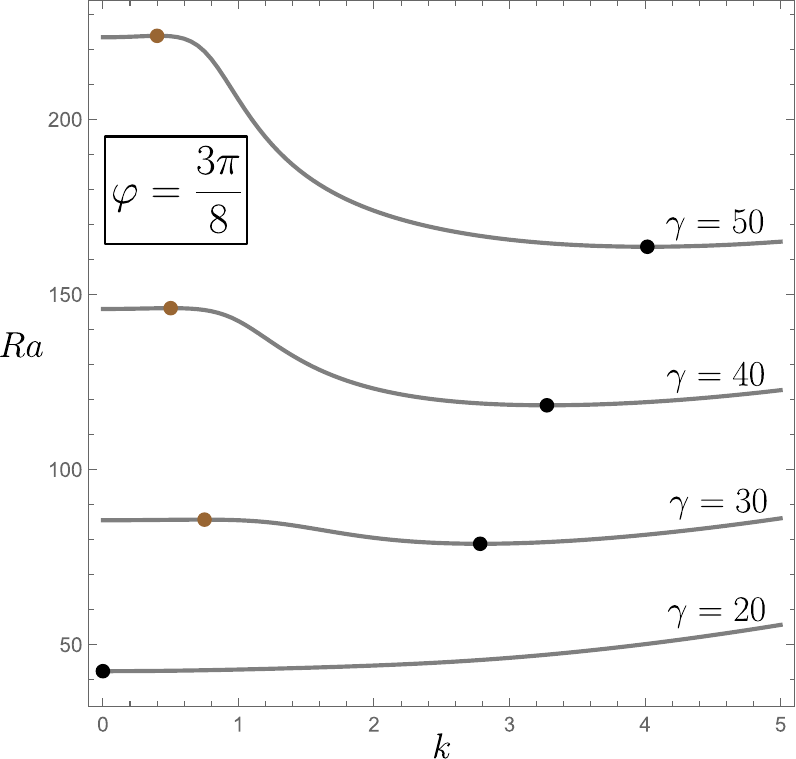}\\[4pt]
\includegraphics[height=7cm]{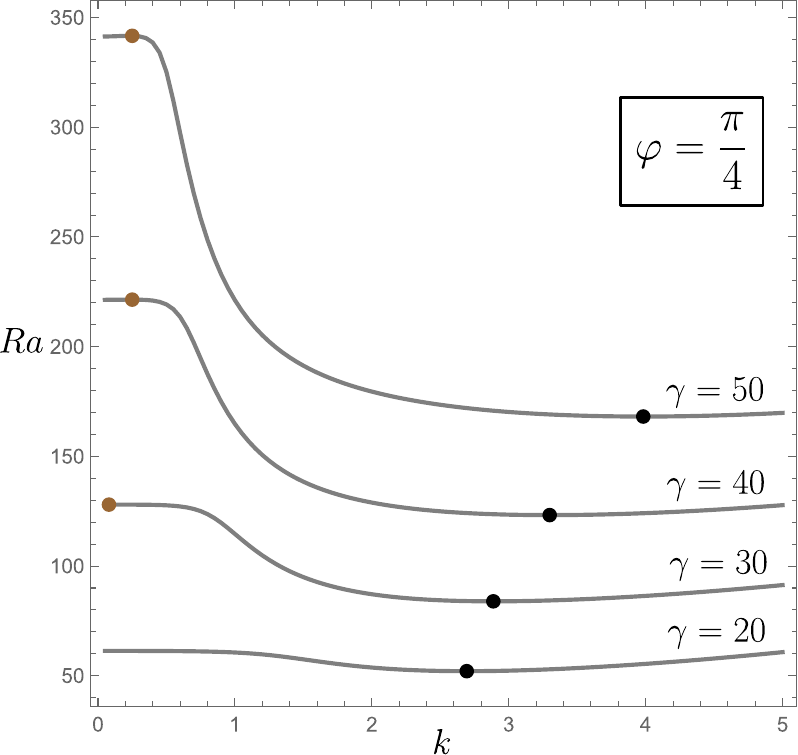}~ \includegraphics[height=7cm]{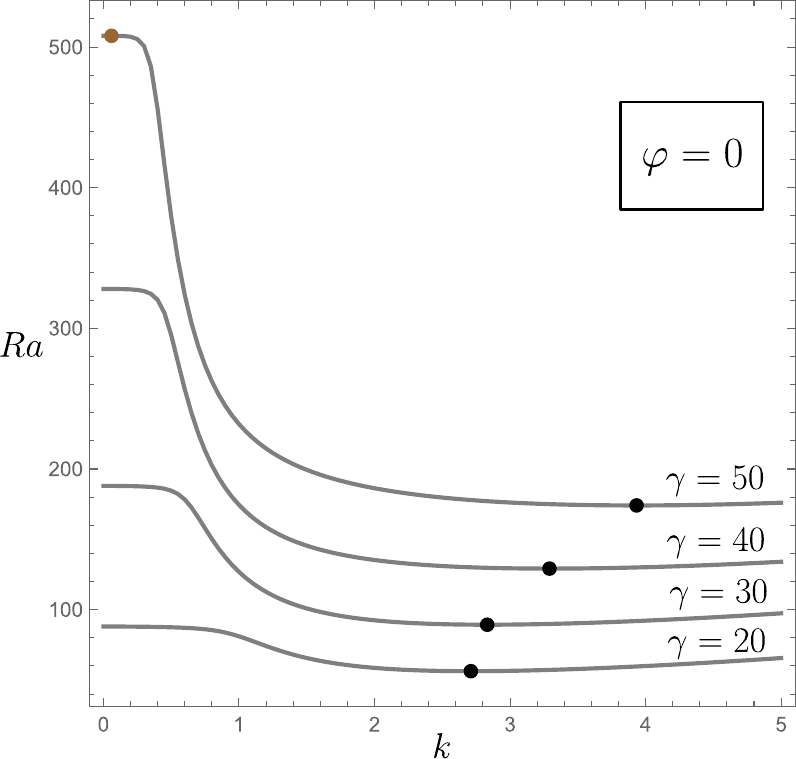}
\caption{\label{fig10}Neutral stability curves in the $(k,Ra)$ plane for $\sigma=-2$ and different values of $\gamma$ and $\varphi$. The black dots denote the critical values $(k_c,Ra_c)$, while the brown dots denote the local maxima of $Ra(k)$ with $k>0$}
\end{figure}

The main symmetry displayed by the eigenvalue problem \eqref{14} is the invariance of the problem under the transformation
\eq{24}{
z \to 1 - z \qc Ra \to - \sigma Ra \qc \sigma \to \frac{1}{\sigma} \qc \gamma \to - \gamma \qc \varphi \to \varphi \qc k \to k \qc \eta \to \eta \qc \tilde{\omega} \to \tilde{\omega}.
}
Such a symmetry justifies the restriction of the parametric domain for the instability analysis to
\eq{25}{
Ra \ge 0 \qc \sigma \in \mathbb{R} \qc \gamma \in \mathbb{R} \qc 0 \le \varphi \le \frac{\pi}{2},
}
where $\mathbb{R}$ denotes the set of real numbers.
There is a further argument showing that reversing the sign of $\gamma$ is unimportant as far as $\qty( Ra, \sigma, |\gamma|, \varphi )$ are kept fixed. Indeed, under this condition, a sign change of $\gamma$ is equivalent to a sign change of $Pe$, as shown by \eqref{7}, which means reversing the direction of the basic flow. One can easily conclude that the orientation of the basic flow, $\vb{n}$ or $-\vb{n}$, is physically insignificant to the instability analysis provided that the inclination of the basic flow streamlines to the $x$ axis, namely the angle $\varphi$, remains unaltered. 
%Hence, it is not restrictive to base the linear instability analysis governed by \eqref{14} on the assumption $\gamma > 0$. 
Thus, one can further restrict the parametric domain defined by \eqref{25} to
\eq{26}{
Ra \ge 0 \qc \sigma \in \mathbb{R} \qc \gamma \ge 0  \qc 0 \le \varphi \le \frac{\pi}{2}.
}

\begin{figure}[ht!]
\centering
\includegraphics[height=7cm]{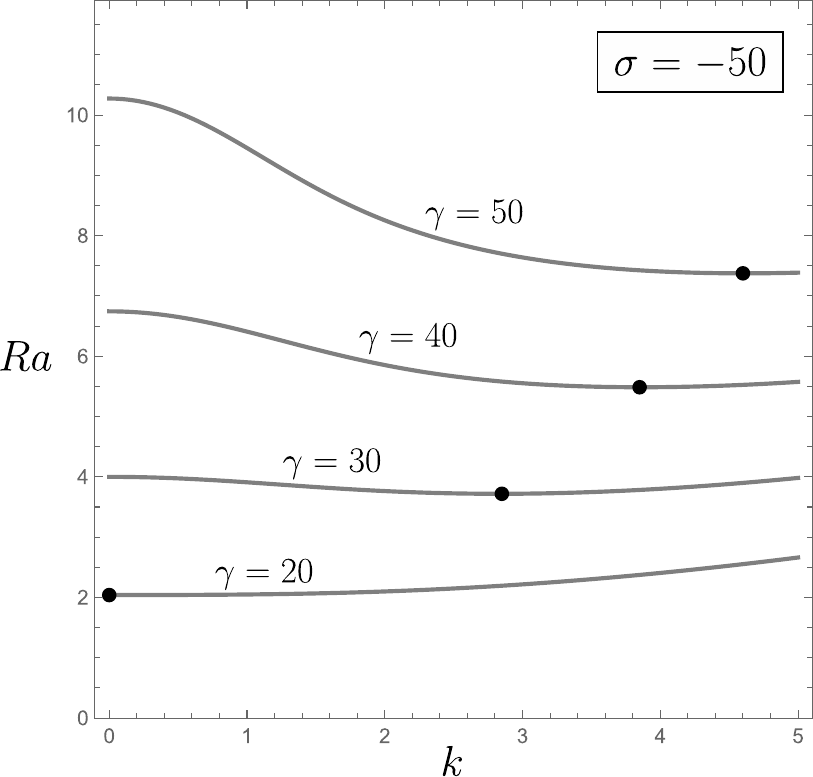}~ \includegraphics[height=7cm]{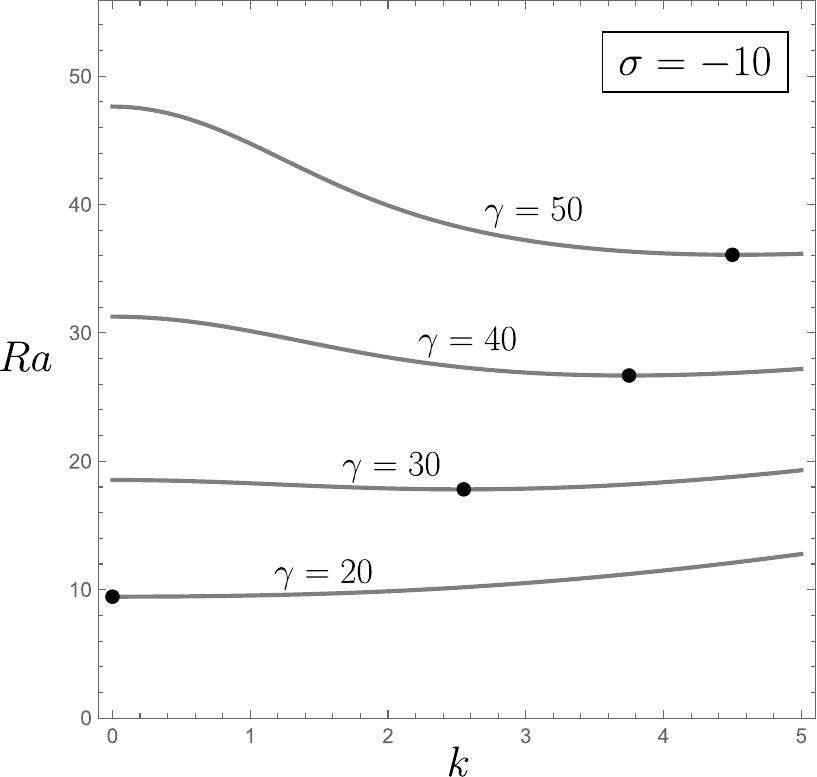}\\[4pt]
\includegraphics[height=7cm]{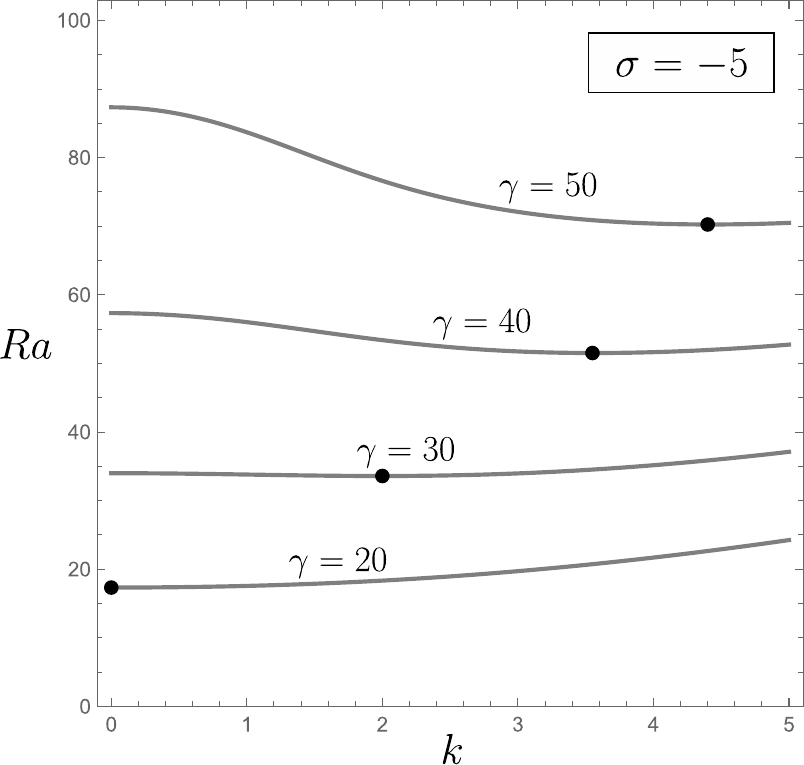}~ \includegraphics[height=7cm]{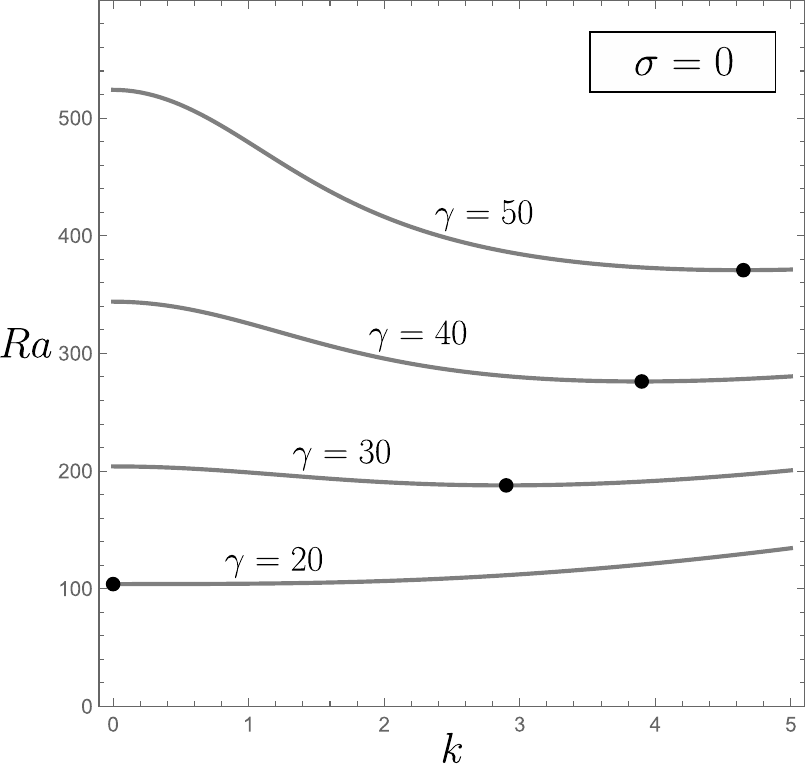}
\caption{\label{fig11}Neutral stability curves in the $(k,Ra)$ plane for longitudinal modes $(\varphi = \pi/2)$ with different values of $\sigma$ and $\gamma$. The black dots denote the critical values $(k_c, Ra_c)$}
\end{figure}

\section{Discussion of the results}
By taking $Ra \ge 0$ with a positive or negative $\sigma$, the neutral stability curves intersect the positive $Ra$ axis only for $\sigma < 1$, as one can easily infer from \eqref{21}. For $\sigma \ge 1$, the neutral stability function $Ra(k)$ turns out to be convex having a singularity when $k \to 0$ and a minimum, $Ra = Ra_c > 0$, for $k = k_c > 0$. Regarding the range $\sigma < 1$, the limit $k \to 0$ of the eigenvalue $Ra$ does not yield the critical Rayleigh number, $Ra_c$, whenever $Ra_2 < 0$. By employing the expression of $Ra_2$ given by \eqref{23}, one obtains the plots displayed in Fig.~\ref{fig4} where the shaded regions identify the condition $Ra_2 > 0$.

One may start the exploration from the asymptotic case where the P\'eclet number is extremely large so that, from \eqref{7}, one has the $\gamma \to 0$ regime. We also point out that the condition $\gamma = 0$ is the only possible one for $\sigma = -1$, independently of the P\'eclet number being large or small. This conclusion is an immediate consequence of the definition of $\gamma$, given by \eqref{7}. 

\subsection{The $\gamma \to 0$ regime}\label{gamma0}
The main feature of the limiting case where $\gamma \to 0$ is the independence of $\varphi$ for the eigenvalue problem \eqref{14}. The independence of $\varphi$ means that every oblique mode is equivalent, in triggering the instability, to transverse and longitudinal modes. This result makes the analysis of this limiting case much simpler than the general case, with the neutral stability curves $Ra(k)$ and the critical values $(k_c, Ra_c, \tilde\omega_c)$ depending only on the heat flux ratio $\sigma$.

A common feature of the neutral stability curves is that, for every $\sigma$ and $k$, the value of $\tilde\omega$ is always zero. Furthermore, Fig~\ref{fig5} shows that the neutral stability curves move upward in the $(k, Ra)$ plane as $\sigma$ increases. The range explored in this figure is from $\sigma = -10$ to $\sigma = 1.5$. When the neutral stability curve moves upward in the $(k, Ra)$ plane, one has a gradual stabilisation of the basic flow, as the region of linear instability is above the neutral stability curve. Furthermore, one may notice that the neutral stability curves, $Ra(k)$, are monotonic increasing starting from $k=0$ up to $\sigma = 0.6$. Indeed, the expression of $Ra_2$ given by \eqref{23} ensures that the function $Ra(k)$ at $k=0$ is convex $(Ra_2> 0)$ provided that
\eq{27}{
\sigma <\frac{1}{19} \left(21-4 \sqrt{5}\right) \approx 0.634512.
}
When the inequality \eqref{27} is not satisfied, then the neutral stability function $Ra(k)$ initially decreases starting from $k=0$, reaches a minimum for $k = k_c >0$ and eventually starts increasing. This behaviour is evident in frame (b) of Fig.~\ref{fig5} for the plots with $0.7 \le \sigma \le 0.9$. All the neutral stability curves with $\sigma \ge 1$ display a singularity at $k=0$ and a minimum $Ra = Ra_c$, for $k= k_c$, which increases with $\sigma$.

Figure~\ref{fig6} shows the plots of the critical values of $k$ and $Ra$ versus $\sigma$ with the brown colour line identifying the case where the onset of convection happens with $k_c=0$, while the black colour line is relative to the case where the onset of convection happens with $k_c>0$, the transition being at $\sigma = 0.634512$. 

\subsection{Perturbation modes with a nonzero $\gamma$}
When $\gamma > 0$, the action of longitudinal, oblique or transverse modes is different, with the angle $\varphi$ significantly affecting the neutral stability curves and the critical values $(k_c, Ra_c, \tilde\omega_c)$. A general feature is that, at neutral stability, $\tilde\omega \ne 0$ for every oblique and transverse modes, {\em i.e.}, when $0 \le \varphi < \pi/2$. 

Figure~\ref{fig7} shows plots of $Ra_c$ versus $\gamma$ with different values of $\sigma$. The brown lines denote those cases where $Ra_c = Ra_0$ and $k_c = 0$, while the black lines are relative to cases where $k_c \ne 0$. The general trends shown by Fig.~\ref{fig7} reveal that $Ra_c$ is a monotonic increasing function of both $\sigma$ and $\gamma$. In fact, an increasing $\sigma$ has a stabilising effect, as the heat flux $q_2$ supplied at the upper boundary increases, so that the thermal buoyancy becomes weaker and weaker. As already pointed out in Section~\ref{gamma0}, as well as in the comments on Fig.~\ref{fig4}, the neutral stability curves in the domain $Ra>0$ do not display an increasing trend for positive values of $k$ close to $0$ when $\sigma$ exceeds the threshold defined by \eqref{27}. Then, for $\sigma$ larger than this threshold, the possibility that $k_c=0$ is ruled out as shown in frame (b) of Fig.~\ref{fig7}.

One may question whether the longitudinal modes considered in Fig.~\ref{fig7} are to be regarded as the most unstable. An argument showing that this is indeed the case is based on the plots reported in Figs.~\ref{fig8} and \ref{fig9}. Such figures display the ratio $Ra_c/Ra_{cL}$ versus $\gamma$, where $Ra_{cL}$ denotes the critical value of $Ra$ evaluated for longitudinal modes. Several different values of $\sigma$ have been considered for the evaluation of $Ra_c/Ra_{cL}$ with three different inclination angles $\varphi$. For every $(\gamma, \sigma)$ included in Figs.~\ref{fig8} and \ref{fig9}, the ratio $Ra_c/Ra_{cL}$ turns out to be a decreasing function of $\varphi$ suggesting that the longitudinal modes, associated with the maximum value, $\varphi = \pi/2$, are the most unstable perturbation modes for the basic state. This consideration means that the transition to linear instability occurs via longitudinal modes, even if there are parametric ranges where $Ra_c$ for oblique or transverse modes is just a few percent larger than that for longitudinal modes. One may reckon that the cases envisaged in Fig.~\ref{fig9} show up such a feature. Both Fig.~\ref{fig8} and Fig.~\ref{fig9} confirm that the limit of $Ra_c/Ra_{cL}$ when $\gamma \to 0$ is $1$. In other words, one retrieves the behaviour in the limiting case $\gamma \to 0$, pointed out in Section~\ref{gamma0}, where the transition to linear instability is unaffected by the inclination angle $\varphi$, so that longitudinal, oblique and transverse modes are equivalent.

We already mentioned that the condition $Ra_2 < 0$ implies that $k_c$ cannot be zero. However, the reverse is untrue, {\em i.e.}, when $Ra_2  > 0$ one may still have the minimum of function $Ra(k)$ for a nonzero wave number. This finding is gathered by comparing Fig.~\ref{fig4} and Fig.~\ref{fig8} in the case $\sigma = -2$. For this case, Fig.~\ref{fig8} shows that $k_c > 0$ when $\gamma$ exceeds a threshold which is approximately $24.0$ for $\varphi = 3\pi/8$, $14.4$ for $\varphi = \pi/4$ and $10.4$ for $\varphi = 0$. On the other hand, Fig.~\ref{fig4} shows that, with $\sigma = -2$, $Ra_2 > 0$ for every $\gamma$ when $\varphi = 3\pi/8$, and only within a limited range of $\gamma$ when $\varphi = \pi/4$ or $0$. The match between the information conveyed by Figs.~\ref{fig4} and \ref{fig8} is found by examining the shape of the neutral stability curves displayed in Fig.~\ref{fig10}. In this figure, the position of the critical point in the $(k, Ra)$ is displayed in each case with a black dot. It is evidenced that there are cases where the neutral stability function $Ra(k)$ is increasing around $k=0$, reaches a maximum (denoted with a brown dot) and then decreases reaching the minimum at $k=k_c > 0$. These are conditions where $Ra_2 > 0$ and, nonetheless, $k_c > 0$ with $Ra_c < Ra_0$. It is worth mentioning that, in every case, the increasing/decreasing trend of $Ra(k)$ around the brown dots, {\em i.e.} the local maxima, is barely perceptible in the plots of Fig.~\ref{fig10}, for scale reasons, while it is utterly evident by inspecting the raw numerical data employed for such plots.

\begin{figure}[t]
\centering
\includegraphics[height=7cm]{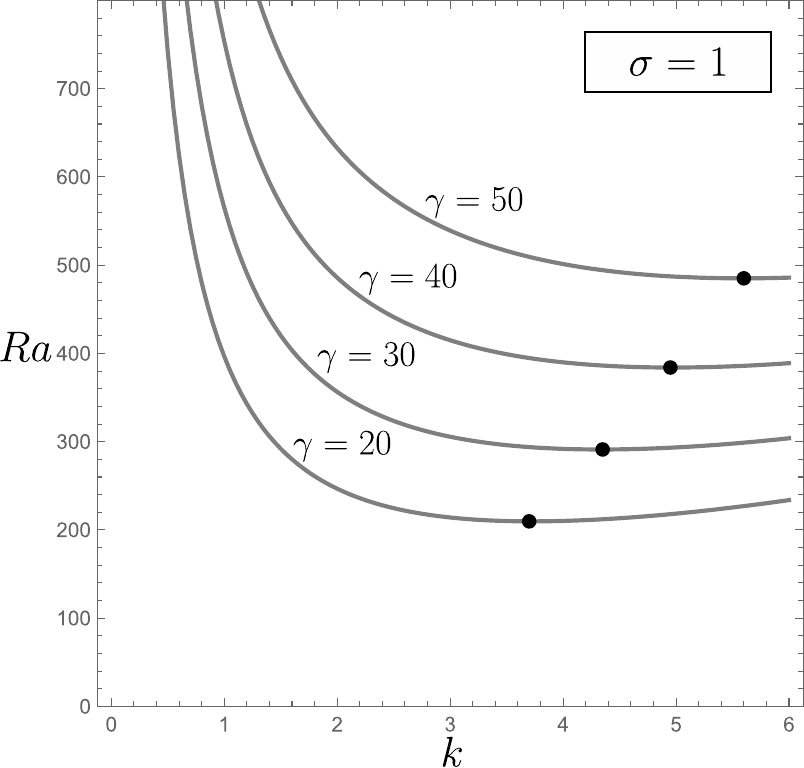}~ \includegraphics[height=7cm]{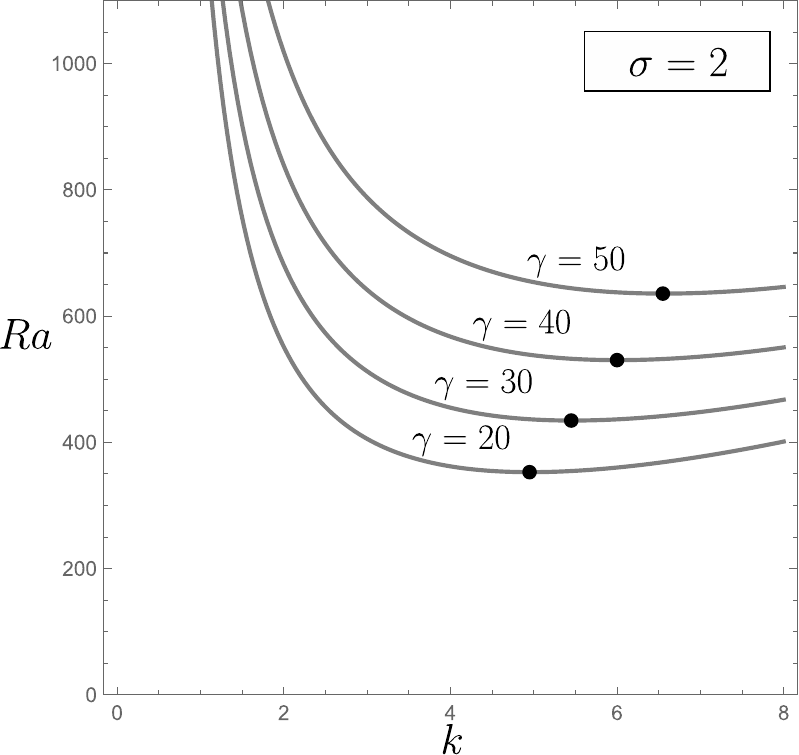}
\caption{\label{fig12}Neutral stability curves in the $(k,Ra)$ plane for longitudinal modes $(\varphi = \pi/2)$ with different values of $\sigma$ and $\gamma$. The black dots denote the critical values $(k_c, Ra_c)$}
\end{figure}

\begin{figure}[ht!]
\centering
\includegraphics[height=7cm]{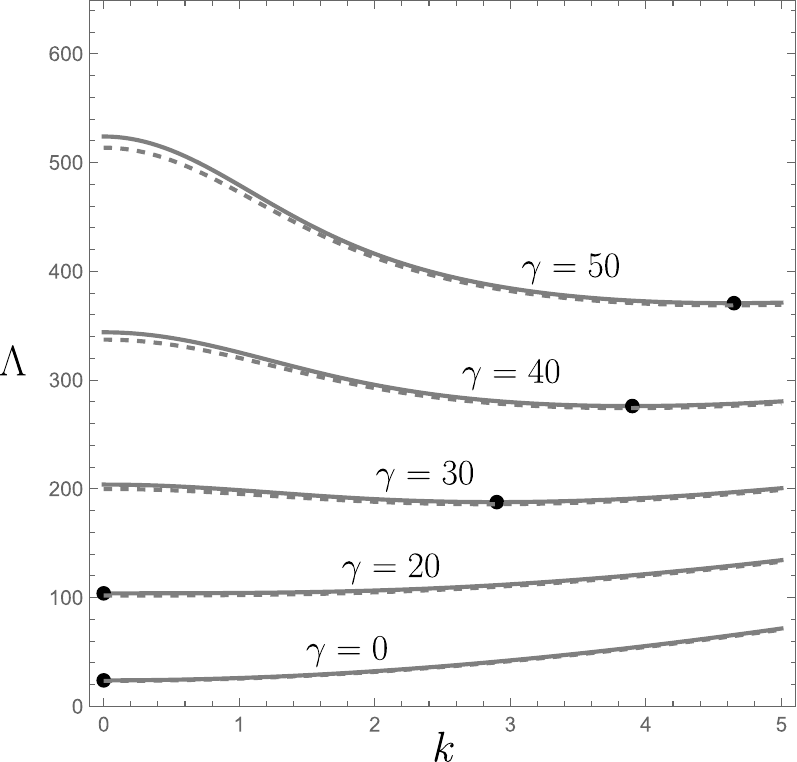}
\caption{\label{fig13}Neutral stability curves in the $(k,\Lambda)$ plane for longitudinal modes $(\varphi = \pi/2)$ having different values of $\gamma$, with either $\sigma \to - \infty$ (solid lines) or $\sigma = - 50$ (dashed lines). The black dots denote the critical values $(k_c, \Lambda_c)$ with $\sigma \to - \infty$}
\end{figure}

\begin{figure}[t]
\centering
\includegraphics[height=7cm]{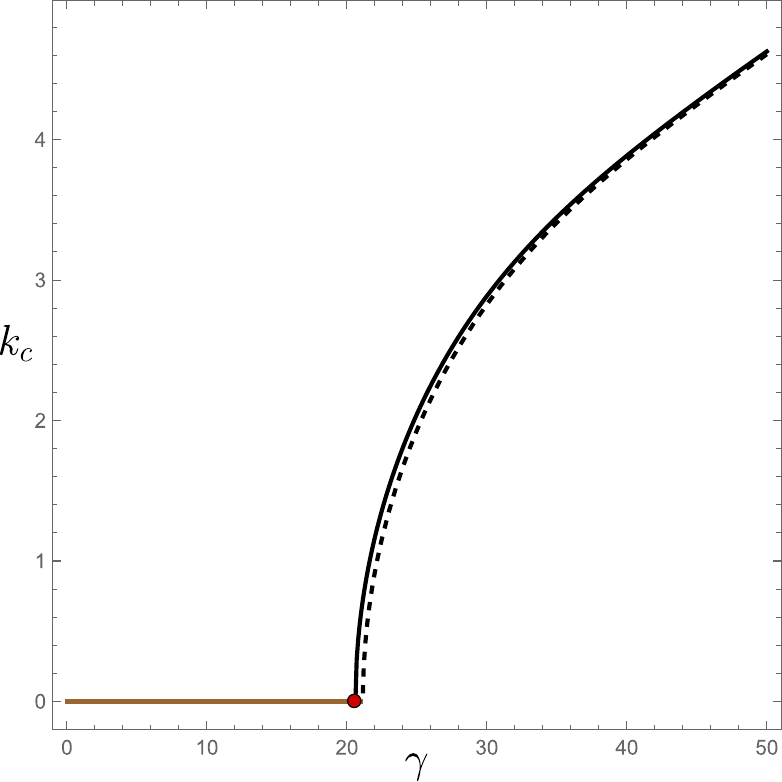}~ \includegraphics[height=7cm]{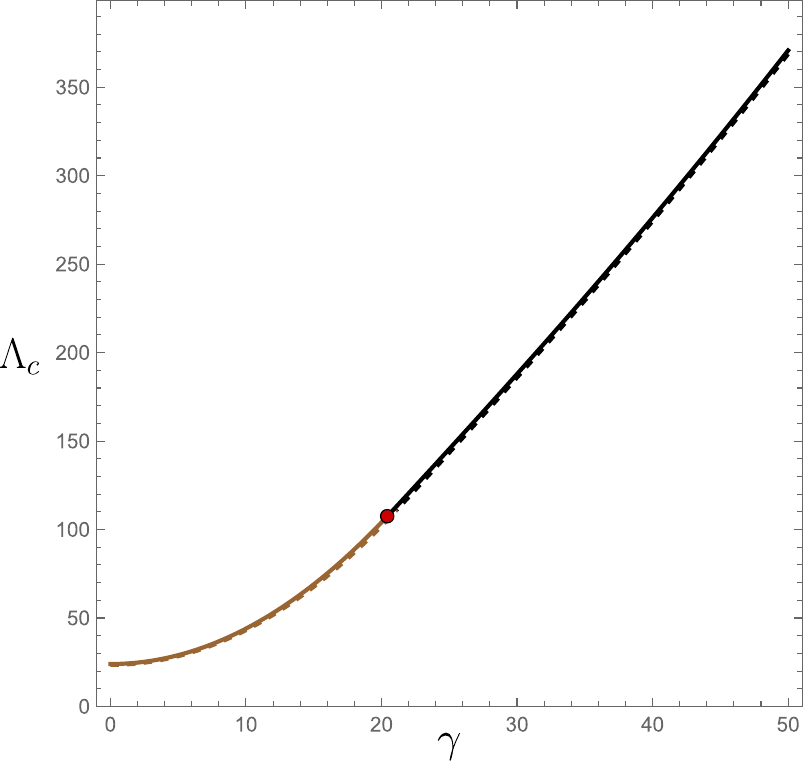}
\caption{\label{fig14}Critical values of $k$ and $\Lambda$ versus $\gamma$ for longitudinal modes $(\varphi = \pi/2)$ with either $\sigma \to - \infty$ (solid lines) or $\sigma = - 50$ (dashed lines). The brown lines are for the range where $k_c=0$, while the black lines are relative to cases where $k_c > 0$}
\end{figure}

Having established, through the arguments presented above, that the transition to linear instability is driven by the longitudinal modes, hereafter,  we will focus on such modes for the neutral stability curves and for the critical values of $k$ and $Ra$. We emphasise again that the neutrally stable longitudinal modes are characterised by $\tilde\omega = 0$. Another important point regards the $Ra_2$ criterion for establishing whether $k_c = 0$ or $k_c > 0$. In fact, for longitudinal modes, $k_c = 0$ if and only if $Ra_2 > 0$. An illustration of the neutral stability curves for longitudinal modes $(\varphi = \pi/2)$ in the $(k,Ra)$ plane is provided in Figs.~\ref{fig11} and \ref{fig12}. While Fig.~\ref{fig11} is relative to cases where $Ra$ at $k=0$ is finite, with $\sigma \le 0$, Fig.~\ref{fig12} displays neutral stability curves where $Ra \to \infty$ when $k \to 0$. Common features for all cases illustrated in these figures are that more stable conditions are found when either $\gamma$ increases or $\sigma$ increases. Strictly speaking, an increase in $\gamma$ means that the streamwise temperature gradient in the basic flow gets larger, while increasing $\sigma$ means reducing the destabilising action of the vertical temperature gradient. In particular, this circumstance suggests that extremely unstable flow conditions occur with $\sigma<0$ when $|\sigma| \gg 1$. 
In order to analyse this asymptotic regime, let us assume that $\sigma \to -\infty$ with $Ra\, \sigma \sim \order{1}$ and $\gamma \sim \order{1}$. Evidently, taking the limit $\sigma \to -\infty$ with $Ra\, \sigma \sim \order{1}$ means that $Ra \to 0$. Thus, one can conveniently introduce a new parameter $\Lambda > 0$, defined as
\eq{28}{
\Lambda = - Ra\, \sigma.
}
By taking the limit $\sigma \to -\infty$ with $\Lambda \sim \order{1}$, $\gamma \sim \order{1}$ and $\varphi = \pi/2$ in the eigenvalue problem \eqref{14}, one obtains
\eq{29}{ 
f'' - k^2\, f+ k\, h=0, \nn
h'' - k^2 \, h -  k   \qty[ \frac{\gamma^2}{2}\, z\,\qty(1-z) - \Lambda \, z] f = 0 ,\nn
z=0,1: \qquad f=0 , \quad  h'=0 ,
}
where we also set $\eta = 0$ as we are interested in the neutral stability condition, and $\tilde\omega = 0$ as expected at neutral stability for longitudinal modes. The asymptotic eigenvalue problem \eqref{29} is solved numerically for a given $\gamma$, so that one can draw the neutral stability curves in the $(k, \Lambda)$ plane. Figure~\ref{fig13} shows a comparison between the neutral stability curves obtained by solving the eigenvalue problem \eqref{29} and those obtained from the data for $\sigma = -50$, already employed for the plots in Fig.~\ref{fig11}. If the curves in Fig.~\ref{fig13} relative to $\sigma \to -\infty$ are identified as solid lines, those for $\sigma = -50$ are displayed as dashed lines. The comparison reveals a fair agreement, showing that one can approximately identify cases with $\sigma < - 50$ with the asymptotic regime $\sigma \to -\infty$. One can also conclude that, in the limit $\sigma \to -\infty$, the critical Rayleigh number drops to zero meaning that the basic flow is unstable for every positive value of $Ra$. Such a result is a straightforward consequence of \eqref{28}. The critical values of $k$ and $\Lambda$ are plotted versus $\gamma$ in Fig.~\ref{fig14}, where the solid lines are relative to $\sigma \to -\infty$, while the dashed lines are for $\sigma = - 50$. This figure also identifies the range where $k_c = 0$ (in brown) and that where $k_c>0$ (in black), the red dots label the transition point, for the limiting case  with $\sigma \to -\infty$, occurring when 
\eq{30}{
\gamma = 2 \sqrt{\frac{10}{23} \left(\sqrt{89551}-55\right)} \approx 20.6103 ,
}
as one can easily infer from the expression \eqref{23} of the coefficient $Ra_2$. Again, Fig.~\ref{fig14} shows the fair agreement between the data for $\sigma \to -\infty$ and those for $\sigma = - 50$.

\begin{table}[t]
\centering
\begin{tabular}{l|ll}
\hline\hline
 $|\gamma|$ & $k_c$  &  $\Lambda_c$  \\ \hline
 0 & 0 &  24  \\
 5 & 0 &  29  \\
 10 & 0  &  44  \\
 20.6103 & 0  &  108.957  \\
 25 & 2.04516  &  145.500  \\
 30 & 2.88622  &  187.871  \\
 40 & 3.88578  &  276.187  \\
 50 & 4.62741  &  370.754  \\
 \hline\hline
\end{tabular}
\caption{\label{tab1}Critical values of $k$ and $\Lambda$ versus $\gamma$ for the limiting case $\sigma \to -\infty$}
\end{table}

If $Ra$ is the Rayleigh number based on the lower wall heat input, $q_1$, the parameter $\Lambda$ defined by \eqref{28}, is the Rayleigh number based on the upper wall heat removal rate, $-q_2$. This is a straightforward consequence of the definition \eqref{5}. Thus, the limit $\sigma \to -\infty$ with $\gamma \sim \order{1}$ can be interpreted, from the physics viewpoint, as one where the lower boundary tends to be thermally insulated $(q_1 \to 0)$ while the upper wall is uniformly cooled $(q_2 < 0)$. This is a practically significant situation. We mention that such an interpretation of the limiting case $\sigma \to -\infty$ should be consistently stated by claiming that $\gamma < 0$, as one may infer from \eqref{7}. However, the sign of $\gamma$ has no effect on the numerical solution of the eigenvalue problem \eqref{29}, as the latter just involves the square of $\gamma$. Therefore, beyond its significance as a mathematical asymptotic case,  the limit $\sigma \to -\infty$ with $\gamma \sim \order{1}$ has a physical interest on its own. For this reason, we report the critical values $k_c$ and $\Lambda_c$ for some given $|\gamma|$ in Table~\ref{tab1}. In fact, we expect no difference between the onset of instability in the case with adiabatic lower boundary and uniform cooling at the upper boundary and that in the case where we have uniform heating from below and an adiabatic upper boundary, namely the case where $\sigma = 0$ \cite{sphaier2014unstable}. From the mathematical viewpoint this circumstance is a consequence of the symmetry defined in \eqref{24}. This is the reason why the critical values of $(k, \Lambda)$ reported in Table~\ref{tab1}, for a given $|\gamma|$, coincide with the critical values of $(k,Ra)$ evaluated for the case $\sigma=0$.   

\section{Conclusions}
A study of the onset of instability in a horizontal porous channel with uniform asymmetric wall heat fluxes at the horizontal boundaries has been carried out. A modal linear theory of perturbation dynamics has been employed to determine the neutral stability condition and the critical values of the wavenumber, $k$, and of the Rayleigh number, $Ra$. The basic stationary state whose instability is investigated involves a basic horizontal flow and a temperature gradient inclined to the vertical. The buoyancy force can trigger the instability for prescribed values of the pair $(\sigma, \gamma)$, where the parameter $\sigma$ is heat flux asymmetry ratio, while the parameter $\gamma$ is the dimensionless horizontal temperature gradient in the basic state. Transition to instability is determined by the Rayleigh number.
The most important results of the linear instability analysis presented in this paper are the following:
\begin{itemize}
\item The longitudinal modes, namely the wavelike modes directed perpendicularly to the basic flow direction, are the most unstable. In other words, the onset of instability is triggered by such modes.
\item The parameter $\sigma$ is stabilising. An increasing value of $\sigma$ from $-\infty$ to $+\infty$ induces an increasing threshold Rayleigh number to linear instability, where such a threshold defines the condition of neutral stability. 
\item The parameter $\gamma$ is also stabilising, where its growth from $0$ to $+\infty$ determines an increasing critical Rayleigh number, $Ra_c$. The most unstable condition, for a fixed $\sigma$, is the limit $\gamma \to 0$ which physically means an infinite flow rate for the basic solution.
\item The limiting case $\sigma \to -\infty$ is the most unstable, for every $\gamma$, and it implies a critical Rayleigh number to instability, $Ra_c$, equal to zero. This means that any basic flow with a positive Rayleigh number is unstable in this limiting case.
\item The onset of the instability may be determined by infinite wavelength modes, {\em i.e.} modes with a zero critical wavenumber, $k_c$, under suitable parametric conditions determined by the pair $(\sigma,  \gamma)$. These conditions have been evaluated analytically through a series expansion in $k$ for the solution of the stability eigenvalue problem.
\end{itemize}
The analysis carried out in this paper is the generalisation of the study published several years ago, regarding the special case where the wall heat fluxes are perfectly symmetric $(\sigma=1)$ \citep{barletta2012thermal}. With respect to that paper, the theoretical framework adopted in this study has an evident difference. In \citet{barletta2012thermal}, the neutral stability curves in the $(k,Ra)$ plane are parametrised by the P\'eclet number $Pe$, while in the present study we have used a parametrisation based on $\gamma$ (for a prescribed $\sigma$).  In practical terms, the parametrisation adopted in \citet{barletta2012thermal} is determined by the basic flow rate, while that adopted in this study is based on the horizontal temperature gradient for the basic state. Both approaches are perfectly legitimate, but the resulting neutral stability curves in the $(k,Ra)$ plane are markedly different. Indeed, the $\gamma$-based approach adopted in this study leads to a mathematical neutral stability function $Ra(k)$ which is single-valued and defines the neutral stability curves in the $(k,Ra)$ plane for every pair $(\sigma, \gamma)$. On the other hand, the $Pe$-based approach leads to neutral stability curves in the $(k,Ra)$ where $Ra(k)$ is not single-valued and, hence, does not define a neutral stability function \citep{barletta2012thermal}. In fact, \citet{barletta2012thermal} found neutral stability curves in the $(k,Ra)$ plane which, for a decreasing $Pe$, tend to assume a closed loop shape. Such closed loops gradually shrink to a point and eventually disappear when $Pe$ attains a minimum below which no linear instability arises \citep{barletta2012thermal}. It's not surprising that different parametrisations lead to seemingly different graphical representations of the same neutral stability condition. The question on which is the best approach to be followed or, at least, the most effective one in the discussion of the instability onset conditions has possibly no good answer. Every different parametrisations of a given stability problem focus on the role played by different physical quantities which, ultimately, reflect the experimental conditions under which the flow dynamics is controlled. In principle, a given experimental apparatus can be designed to fix either the horizontal temperature gradient (the $\gamma$-based approach) or the flow rate (the $Pe$-based approach) in the basic state.

\subsection*{Acknowledgements} 
The work by Antonio Barletta, Michele Celli and Pedro Vayssi\`ere Brand\~ao was supported by Alma Mater Studiorum Universit\`a di Bologna, grant number
RFO-2024.

\bibliography{biblio}

\begin{thebibliography}{23}
\expandafter\ifx\csname natexlab\endcsname\relax\def\natexlab#1{#1}\fi
\providecommand{\bibinfo}[2]{#2}
\ifx\xfnm\relax \def\xfnm[#1]{\unskip,\space#1}\fi
%Type = Book
\bibitem[{Nield and Bejan(2017)}]{NiBe17}
\bibinfo{author}{D.~A. Nield}, \bibinfo{author}{A.~Bejan},
  \bibinfo{title}{Convection in Porous Media}, \bibinfo{publisher}{Springer},
  \bibinfo{address}{New York}, \bibinfo{edition}{5th} edition,
  \bibinfo{year}{2017}.
%Type = Article
\bibitem[{Prats(1966)}]{prats1966effect}
\bibinfo{author}{M.~Prats},
\newblock \bibinfo{title}{The effect of horizontal fluid flow on thermally
  induced convection currents in porous mediums},
\newblock \bibinfo{journal}{Journal of Geophysical Research}
  \bibinfo{volume}{71} (\bibinfo{year}{1966}) \bibinfo{pages}{4835--4838}.
%Type = Article
\bibitem[{Dufour and Neel(1998)}]{1010631869741}
\bibinfo{author}{F.~Dufour}, \bibinfo{author}{M.-C. Neel},
\newblock \bibinfo{title}{Numerical study of instability in a horizontal porous
  channel with bottom heating and forced horizontal flow},
\newblock \bibinfo{journal}{Physics of Fluids} \bibinfo{volume}{10}
  (\bibinfo{year}{1998}) \bibinfo{pages}{2198--2207}.
%Type = Article
\bibitem[{Chung et~al.(2002)Chung, Park, Choi, and Yoon}]{chung2002onset}
\bibinfo{author}{T.~J. Chung}, \bibinfo{author}{J.~H. Park},
  \bibinfo{author}{C.~K. Choi}, \bibinfo{author}{D.-Y. Yoon},
\newblock \bibinfo{title}{The onset of vortex instability in laminar forced
  convection flow through a horizontal porous channel},
\newblock \bibinfo{journal}{International Journal of Heat and Mass Transfer}
  \bibinfo{volume}{45} (\bibinfo{year}{2002}) \bibinfo{pages}{3061--3064}.
%Type = Article
\bibitem[{Ouarzazi et~al.(2008)Ouarzazi, Mejni, Delache, and
  Labrosse}]{ouarzazi2008nonlinear}
\bibinfo{author}{M.~N. Ouarzazi}, \bibinfo{author}{F.~Mejni},
  \bibinfo{author}{A.~Delache}, \bibinfo{author}{G.~Labrosse},
\newblock \bibinfo{title}{Nonlinear global modes in inhomogeneous mixed
  convection flows in porous media},
\newblock \bibinfo{journal}{Journal of Fluid Mechanics} \bibinfo{volume}{595}
  (\bibinfo{year}{2008}) \bibinfo{pages}{367--377}.
%Type = Article
\bibitem[{Delache and Ouarzazi(2008)}]{delache2008weakly}
\bibinfo{author}{A.~Delache}, \bibinfo{author}{M.~Ouarzazi},
\newblock \bibinfo{title}{Weakly nonlinear interaction of mixed convection
  patterns in porous media heated from below},
\newblock \bibinfo{journal}{International Journal of Thermal Sciences}
  \bibinfo{volume}{47} (\bibinfo{year}{2008}) \bibinfo{pages}{709--722}.
%Type = Article
\bibitem[{Chung et~al.(2010)Chung, Choi, Yoon, and Kim}]{chung2010onset}
\bibinfo{author}{T.~J. Chung}, \bibinfo{author}{C.~K. Choi},
  \bibinfo{author}{D.-Y. Yoon}, \bibinfo{author}{M.~C. Kim},
\newblock \bibinfo{title}{Onset of buoyancy-driven motion with laminar forced
  convection flows in a horizontal porous channel},
\newblock \bibinfo{journal}{International Journal of Heat and Mass Transfer}
  \bibinfo{volume}{53} (\bibinfo{year}{2010}) \bibinfo{pages}{5139--5146}.
%Type = Article
\bibitem[{Barletta(2012)}]{barletta2012thermal}
\bibinfo{author}{A.~Barletta},
\newblock \bibinfo{title}{Thermal instability in a horizontal porous channel
  with horizontal through flow and symmetric wall heat fluxes},
\newblock \bibinfo{journal}{Transport in Porous Media} \bibinfo{volume}{92}
  (\bibinfo{year}{2012}) \bibinfo{pages}{419--437}.
%Type = Article
\bibitem[{Barletta et~al.(2014)Barletta, Celli, and
  Kuznetsov}]{barletta2014convective}
\bibinfo{author}{A.~Barletta}, \bibinfo{author}{M.~Celli},
  \bibinfo{author}{A.~V. Kuznetsov},
\newblock \bibinfo{title}{Convective instability of the {D}arcy flow in a
  horizontal layer with symmetric wall heat fluxes and local thermal
  nonequilibrium},
\newblock \bibinfo{journal}{ASME Journal of Heat Transfer}
  \bibinfo{volume}{136} (\bibinfo{year}{2014}) \bibinfo{pages}{012601}.
%Type = Article
\bibitem[{Sphaier and Barletta(2014)}]{sphaier2014unstable}
\bibinfo{author}{L.~A. Sphaier}, \bibinfo{author}{A.~Barletta},
\newblock \bibinfo{title}{Unstable mixed convection in a heated horizontal
  porous channel},
\newblock \bibinfo{journal}{International Journal of Thermal Sciences}
  \bibinfo{volume}{78} (\bibinfo{year}{2014}) \bibinfo{pages}{77--89}.
%Type = Article
\bibitem[{Dubey and Murthy(2018)}]{dubey2018onset}
\bibinfo{author}{R.~Dubey}, \bibinfo{author}{P.~V. S.~N. Murthy},
\newblock \bibinfo{title}{The onset of convective instability of horizontal
  throughflow in a porous layer with inclined thermal and solutal gradients},
\newblock \bibinfo{journal}{Physics of Fluids} \bibinfo{volume}{30}
  (\bibinfo{year}{2018}) \bibinfo{pages}{074104}.
%Type = Article
\bibitem[{Barletta and Rees(2019)}]{barletta2019unstable}
\bibinfo{author}{A.~Barletta}, \bibinfo{author}{D.~A.~S. Rees},
\newblock \bibinfo{title}{Unstable mixed convection flow in a horizontal porous
  channel with uniform wall heat flux},
\newblock \bibinfo{journal}{Transport in Porous Media} \bibinfo{volume}{129}
  (\bibinfo{year}{2019}) \bibinfo{pages}{385--402}.
%Type = Article
\bibitem[{Barletta et~al.(2024)Barletta, Celli, Brand{\~a}o, Lazzari, and
  Ghedini}]{barletta2024linearly}
\bibinfo{author}{A.~Barletta}, \bibinfo{author}{M.~Celli},
  \bibinfo{author}{P.~V. Brand{\~a}o}, \bibinfo{author}{S.~Lazzari},
  \bibinfo{author}{E.~Ghedini},
\newblock \bibinfo{title}{Linearly unstable forced and free flow in an
  anisotropic porous channel},
\newblock \bibinfo{journal}{International Journal of Heat and Mass Transfer}
  \bibinfo{volume}{235} (\bibinfo{year}{2024}) \bibinfo{pages}{126155}.
%Type = Article
\bibitem[{Dubey and Chetteti(2024)}]{dubey2024influence}
\bibinfo{author}{R.~Dubey}, \bibinfo{author}{R.~Chetteti},
\newblock \bibinfo{title}{The influence of thermal dispersion on the initiation
  of convective instability in {P}rats flow through a low permeability porous
  medium},
\newblock \bibinfo{journal}{Physics of Fluids} \bibinfo{volume}{36}
  (\bibinfo{year}{2024}) \bibinfo{pages}{014104}.
%Type = Book
\bibitem[{Straughan(2013)}]{straughan2013energy}
\bibinfo{author}{B.~Straughan}, \bibinfo{title}{The Energy Method, Stability,
  and Nonlinear Convection}, \bibinfo{publisher}{Springer},
  \bibinfo{year}{2013}.
%Type = Book
\bibitem[{Barletta(2019)}]{barletta2019routes}
\bibinfo{author}{A.~Barletta}, \bibinfo{title}{Routes to Absolute Instability
  in Porous Media}, \bibinfo{publisher}{Springer}, \bibinfo{year}{2019}.
%Type = Article
\bibitem[{Buonomo et~al.(2020)Buonomo, di~Pasqua, Manca, and
  Nardini}]{buonomo2020evaluation}
\bibinfo{author}{B.~Buonomo}, \bibinfo{author}{A.~di~Pasqua},
  \bibinfo{author}{O.~Manca}, \bibinfo{author}{S.~Nardini},
\newblock \bibinfo{title}{Evaluation of thermal and fluid dynamic performance
  parameters in aluminum foam compact heat exchangers},
\newblock \bibinfo{journal}{Applied Thermal Engineering} \bibinfo{volume}{176}
  (\bibinfo{year}{2020}) \bibinfo{pages}{115456}.
%Type = Article
\bibitem[{Bianco et~al.(2021)Bianco, Buonomo, di~Pasqua, and
  Manca}]{bianco2021heat}
\bibinfo{author}{V.~Bianco}, \bibinfo{author}{B.~Buonomo},
  \bibinfo{author}{A.~di~Pasqua}, \bibinfo{author}{O.~Manca},
\newblock \bibinfo{title}{Heat transfer enhancement of laminar impinging slot
  jets by nanofluids and metal foams},
\newblock \bibinfo{journal}{Thermal Science and Engineering Progress}
  \bibinfo{volume}{22} (\bibinfo{year}{2021}) \bibinfo{pages}{100860}.
%Type = Article
\bibitem[{Rajagopal et~al.(1996)Rajagopal, Ruzicka, and
  Srinivasa}]{rajagopal1996oberbeck}
\bibinfo{author}{K.~R. Rajagopal}, \bibinfo{author}{M.~Ruzicka},
  \bibinfo{author}{A.~R. Srinivasa},
\newblock \bibinfo{title}{On the {O}berbeck-{B}oussinesq approximation},
\newblock \bibinfo{journal}{Mathematical Models and Methods in Applied
  Sciences} \bibinfo{volume}{6} (\bibinfo{year}{1996})
  \bibinfo{pages}{1157--1167}.
%Type = Article
\bibitem[{Barletta(2022)}]{barletta2022boussinesq}
\bibinfo{author}{A.~Barletta},
\newblock \bibinfo{title}{The {B}oussinesq approximation for buoyant flows},
\newblock \bibinfo{journal}{Mechanics Research Communications}
  \bibinfo{volume}{124} (\bibinfo{year}{2022}) \bibinfo{pages}{103939}.
%Type = Article
\bibitem[{Brand\~ao et~al.(2021)Brand\~ao, Barletta, Celli, Alves, and
  Rees}]{brandao2021stability}
\bibinfo{author}{P.~V. Brand\~ao}, \bibinfo{author}{A.~Barletta},
  \bibinfo{author}{M.~Celli}, \bibinfo{author}{L.~S. d.~B. Alves},
  \bibinfo{author}{D.~A.~S. Rees},
\newblock \bibinfo{title}{On the stability of the isoflux {D}arcy-{B\'e}nard
  problem with a generalised basic state},
\newblock \bibinfo{journal}{International Journal of Heat and Mass Transfer}
  \bibinfo{volume}{177} (\bibinfo{year}{2021}) \bibinfo{pages}{121538}.
%Type = Article
\bibitem[{Turkyilmazoglu and Siddiqui(2023)}]{turkyilmazoglu2023instability}
\bibinfo{author}{M.~Turkyilmazoglu}, \bibinfo{author}{A.~A. Siddiqui},
\newblock \bibinfo{title}{The instability onset of generalized isoflux mean
  flow using {B}rinkman-{D}arcy-{B\'e}nard model in a fluid saturated porous
  channel},
\newblock \bibinfo{journal}{International Journal of Thermal Sciences}
  \bibinfo{volume}{188} (\bibinfo{year}{2023}) \bibinfo{pages}{108249}.
%Type = Misc
\bibitem[{Mat(2024)}]{Mathematica}
\bibinfo{title}{Wolfram {R}esearch{,} {I}nc., {M}athematica, {V}ersion 14.2},
  \bibinfo{howpublished}{{\tt https://www.wolfram.com/mathematica}},
  \bibinfo{year}{2024}. \bibinfo{note}{Champaign, IL}.

\end{thebibliography}

\end{document}